\documentclass{aastex}
\usepackage{emulateapj5}
\newcommand{\be}{\begin{equation}}
\newcommand{\ee}{\end{equation}}
\newcommand{\bea}{\begin{eqnarray}}
\newcommand{\eea}{\end{eqnarray}}

\shorttitle{Growth of Magnetic Field}
\shortauthors{Cho et al.}

\begin{document}
\title{Growth of Magnetic Fields Induced by Turbulent Motions} 
\author{Jungyeon Cho\altaffilmark{1}, Ethan T. Vishniac\altaffilmark{2}, 
        Andrey Beresnyak\altaffilmark{3}, A. Lazarian\altaffilmark{3}, 
        and Dongsu Ryu\altaffilmark{1}}  

\altaffiltext{1}{Dept. of Astronomy and Space Science,
       Chungnam National University, Daejeon, Korea}
\altaffiltext{2}{Dept. of Physics and Astronomy,
       McMaster University, Hamilton, Ontario L85 4M1, Canada}
\altaffiltext{3}{Dept. of Astronomy, University of Wisconsin,
       Madison, WI53705}




\begin{abstract}
We present numerical simulations of driven magnetohydrodynamic (MHD)
turbulence with weak/moderate imposed magnetic fields.
The main goal is to clarify 
dynamics of magnetic field growth.
We also investigate the effects 
of the imposed magnetic fields on
the MHD turbulence, including, as a limit,
the case of zero external field. 
Our findings are as follows.
First, when we start off simulations with weak mean magnetic field only
(or with small scale random field with zero imposed field),
we observe that there is a stage at which magnetic energy
density grows linearly with time.
Runs with different numerical resolutions and/or 
different simulation parameters show consistent results for
the growth rate at the linear stage.
Second, we find that, when the strength of the external field increases,
the equilibrium kinetic energy density drops by roughly the product of the rms 
velocity and the
strength of the external field. 
The equilibrium magnetic energy density rises by 
roughly the same amount.   
Third, when the external magnetic field 
is not very strong (say, less than $\sim 0.2$ times the rms velocity
when measured in the units of Alfven speed),
the turbulence at large scales remains statistically isotropic, i.e. there
is no apparent {\it global} anisotropy of order $B_0/v$.
We discuss implications of our results on astrophysical fluids.
\end{abstract}

\section{Introduction}
 Most astrophysical fluids are magnetized.
 Magnetic field in an astrophysical system can be divided into two components:
 large-scale regular and small-scale random components. 
 The generation of magnetic field may involve with two separate issues: 
generation of
 the large-scale regular field (or mean field) 
and generation of small-scale random field.

The generation or growth of large-scale regular fields is an important topic
in astrophysics. However, in this paper we assume fixed large-scale regular 
fields are
already present and we only investigate how small-scale random fields are 
generated from the imposed large-scale fields.
That is,
we investigate the growth of magnetic energy in the presence of fixed mean fields. 
Therefore, the generation/growth of large-scale regular 
field itself is not a topic of
this paper.
Readers interested in the topic may refer to 
mean-field dynamo theories (see Moffatt 1978; Parker 1979; Krause \& Radler 1980;
Brandenburg \& Subramanian 2005).  
The type of magnetic energy growth we deal with in this paper
is so-called small-scale {\it turbulence dynamo}\footnote{
 In this paper, 
 turbulence dynamo means {\it not} growth of mean field itself 
 {\it but}  
 generation of random field on scales
 similar to or smaller than the driving scale of turbulence, 
 which is sometimes
 referred to as fluctuation dynamo 
 (see, for example, Schekochihin et al. 2007;
 Subramanian 2008). 
 In all our simulations the strengths of the mean fields do not change
 with time.
 In this paper, by `amplification of magnetic field' or
 `growth of magnetic field',
 we actually mean amplification of 
 fluctuating field(s) on scales similar to or smaller than the driving scale.
 Therefore, the small-scale turbulence dynamo is different from
 mean-field dynamo, which deals with growth of mean field itself.
}, the origin of which can be
traced back to the papers by Batchelor (1950) and Kazantsev (1968). 
We also study how properties of magnetohydrodynamic (MHD) 
turbulence (e.g. energy densities, power spectra, global
anisotropy) change as the strength of the mean field changes.

When we introduce a mean field in a turbulent medium, the mean field
interacts with turbulent motions. There are two distinct types of interaction, depending on the
strength of the mean magnetic field.
When the large-scale regular magnetic field is weak, turbulent motions stretch the magnetic field lines and, as a result, 
the magnetic energy density increases. 
MHD turbulence near the scale of the largest energy-containing eddies (i.e, the outer scale or the energy injection scale)
will be more or less like ordinary hydrodynamic turbulence with small magnetic back reaction. 

MHD turbulence in intracluster medium and intergalactic medium may fall in this type of
turbulence.
The origin of the seed magnetic fields is still uncertain.
However, whatever the origin is, turbulence motions can
produce small-scale field through the stretching of the seed fields.
In this regime, the 
growth timescale will be of great importance.
If the growth time scale is shorter than the age of the universe in a system,
we expect that the system is strongly magnetized in the present time.
Otherwise, we expect that the system is weakly magnetized.

On the other hand, when the imposed mean magnetic field is strong in the sense that the turbulent eddy turnover rate
at the large scale (i.e. $L/v$) is  slower than the Alfvenic rate of the same scale 
(i.e. $L/B_0$)\footnote{In what follows we measure the magnetic field in the units of Alfven velocity $V_A$.}, 
the resulting turbulence can be described through the nonlinear interaction of waves. The turbulence can be either {\it strong}, meaning that the cascading happens within one eddy turnover time or {\it weak}, meaning that the cascading takes more than one eddy turnover time. A classical study of Iroshnikov (1963) and Kraichnan (1965) presents an example of weak MHD turbulence. This is a hypothetical {\it isotropic} MHD turbulence, while we know by now that the actual MHD turbulence is anisotropic (Shebalin et al 1983; Higdon 1986;
Goldreich \& Sridhar 1995; Matthaeus et al. 1998;
see also Biskamp 2003 and references therein).
An exact analytical treatment of weak MHD turbulence can be found in
Galtier et al.~(2000).
 A successful\footnote{The Goldreich \& Sridhar model was successfully tested in incompressible 3D MHD simulations in Cho \& Vishniac (2000b), Maron \& Goldreich (2001), Cho, Lazarian \& Vishniac (2002), as well as in compressible 3D MHD
 simulations in Cho \& Lazarian (2002, 2003). While the particular points of the model, i.e. the particular slope of the spectrum
 are still the subject of debates (see Maron \& Goldreich 2001; Boldyrev 2005, 2006; Beresnyak \& Lazarian 2006; Mason, Cattaneo \& Boldyrev 2006), the corner stone of the model, which is the critical balance, stays untouched.}  model of the strong MHD turbulence was presented in  Goldreich \& Sridhar (1995).

In this paper we deal with the former case: turbulence with weak imposed magnetic field.
In the interstellar medium community, this type of turbulence is called
 super-Alfvenic turbulence, which is favored by some researchers (see Padoan \& Nordlund 1999; Padoan et al. 2004) as a model of turbulence
 in molecular clouds. In any case, such turbulence is expected to be present in any system with magnetic field below
 the equipartition value, which gets subject to intensive driving.  As the turbulence kinetic energy decreases with the scale, 
 we expect
 the Goldreich-Sridhar (1995) model to be valid at some small scale when the equipartition is reached, while at larger scales we expect hydrodynamic motions to increase the energy of magnetic field.

It is a common knowledge that the effects of mean magnetic field have important astrophysical implications.
In recent years, interest on the turbulent processes in tangled magnetic field has been growing.
Relevant astrophysical problems include thermal diffusion in the intracluster medium (see Chandran \& Cowley 1998;
Narayan \& Medvedev 2001; Cho et al. 2003; Lazarian 2006), cosmic ray propagation (see  Cassano \& Brunetti 2005; Lazarian \& Beresnyak 2006; Brunetti \& Lazarian 2007) as well as star formation (Padoan et al. 2004; Li \& Nakamura 2004; Vazquez-semadeni, Kim \& Ballesteros-Paredes 2005).

Cho \& Vishniac (2000a) numerically showed that magnetic energy 
grows until the magnetic energy density gets comparable to the kinetic energy density
(see also Kulsrud \& Anderson 1992; Kulsrud et al.~1997). 
In this paper, we present more comprehensive studies on the topic.
Other aspects of the magnetic field generation also require further studies.
For example, Haugen \& Brandenburg (2004) discussed spectral change of
MHD turbulence by mean field and showed that imposed magnetic field
lowers the spectral magnetic energy in the inertial range.
Mac Low (1999) demonstrated that mean magnetic field produces anisotropic
structures along the mean field direction in strongly compressible 
MHD turbulence.
 Lee et al. (2003) discussed the behaviors of energy densities and spectral shapes for 
three different cases (very weak, weak, and strong mean magnetic field cases) but only for 2-dimensional  MHD turbulence.
They showed the flow character in very weak field classes is similar to
that of hydrodynamic turbulence, while the strong field cases show spectra shallower than the hydrodynamic one.
Schekochihin et al. (2007) discussed the effects of mean fields
for different values of magnetic Prandtl numbers.
In all the papers above, the increase of magnetic field energy was noticed.
In this paper, we present a comprehensive study on the effect of mean magnetic field on 3-dimensional MHD turbulence and turbulence dynamo.

Another issue that requires clarification is the effect of mean magnetic field on the decay of MHD turbulence.
Stone, Ostriker, \& Gammie (1998) and Mac Low et al. (1998) numerically showed
that damping time-scales of compressible MHD turbulence are comparable
to the large-scale eddy turnover time (see McKee \& Ostriker 2007 for
a collection of related results). Incompressible MHD turbulence also decays fast (see
Cho et al. 2002). While the earlier works were mostly focused on two extreme limits - 
zero (see, for example, Biskamp \& Muller 2000) and
strong mean-field limits.
 In what follows we discuss how the decay time-scale changes as 
the mean field strength changes.

We will first consider the regime of very weak
mean field. In this regime, we will mainly investigate 
the growth rate of magnetic field.
Then, we will consider the effect of intermediate mean fields.
We will investigate how magnetic and kinetic energy
 densities, anisotropy, and growth rate of magnetic field change
with the increase of the strength of the mean field.
In this paper, we deal with incompressible MHD turbulence.
We describe our numerical methods in \S 2, present our results for the very weak mean-field regime in \S 3 and
discuss the effects of the mean field in \S 4. We compare MHD and hydrodynamic turbulence in \S 5 and
we give discussion in \S 6.
We provide our conclusions in \S 7.

\section{Numerical Methods}
We have used a pseudospectral code to solve the 
incompressible MHD equations in a periodic box of size $2\pi$:
\begin{equation}
\frac{\partial {\bf v} }{\partial t} = (\nabla \times {\bf v}) \times {\bf v}
      -(\nabla \times {\bf B})
        \times {\bf B} + \nu \nabla^{2} {\bf v} + {\bf f} + \nabla P' ,
        \label{veq}
\end{equation}
\begin{equation}
\frac{\partial {\bf B}}{\partial t}= 
     \nabla \times ({\bf v} \times{\bf B}) + \eta \nabla^{2} {\bf B} ,
     \label{beq}
\end{equation}
\be
      \nabla \cdot {\bf v} =\nabla \cdot {\bf B}= 0,
\ee
where $\bf{f}$ is random driving force,
$P'\equiv P + {\bf v}\cdot {\bf v}/2$, ${\bf v}$ is the velocity,
and ${\bf B}$ is magnetic field divided by $(4\pi \rho)^{1/2}$.
We use 21 forcing components with $2\leq k \leq \sqrt{12}$.
Each forcing component has correlation time of one.
The peak of energy injection occurs at $k\approx 2.5 $.
The amplitudes of the forcing components are tuned to ensure $v \approx 1$
for the hydrodynamic simulation with $\nu=0.0074$.
Therefore, one eddy turnover time, $\sim L/v$, is approximately 2.5 time units.
In this representation, ${\bf v}$ can be viewed as the velocity 
measured in units of the r.m.s. velocity, v,
of the system and ${\bf B}$ as the Alfven speed in the same units.
Other variables have their usual meaning.
The magnetic field consists of the uniform background field and the 
fluctuating field: ${\bf B}= {\bf B}_0 + {\bf b}$.
The Alfv\'{e}n velocity of
the background field, $B_0$,
varies from 0 to 1.
Through out the paper, we consider only cases where viscosity is
equal to magnetic diffusion\footnote{
       In this paper, our goal is to 
       study small-scale turbulence dynamo properties in the asymptotic limit of 
       very small viscosity and very small magnetic diffusion.
     Unit magnetic Prandtl number ($\nu=\eta$) will be a good approximation 
     in this limit.
     To support this claim, we ran a simulation which is very similar to
     $256H3-B_010^{-3}$ but the magnetic Prandtl number ($=\nu/\eta$) is 0.01.
     The qualitative behavior of time evolution (not shown in this paper) 
     is similar to that of the unit magnetic
     Prandtl number case: magnetic energy density grows initially and saturates later
     when magnetic energy density becomes comparable to the kinetic one.
     The average values of $v^2$ and $b^2$ are $\sim 0.78$ and
     $\sim 0.25$, respectively. Average is taken over the time interval of (50,110).
     The magnetic energy density is smaller than that of $256H3-B_010^{-3}$
     (see Table 1)
     because magnetic dissipation occurs at smaller wavenumbers.
}:
\be
  \nu = \eta.
\ee
In pseudospectral methods, we calculate the temporal evolution of
the equations (\ref{veq}) and (\ref{beq}) in Fourier space.
To obtain the Fourier components of nonlinear terms, we first calculate
them in real space, and transform back into Fourier space.
We use exactly same forcing terms
for all simulations.
The average kinetic helicity in these simulations is negative. 

We use an appropriate projection operator to calculate 
$\nabla P'$ term in
{}Fourier space and also to enforce divergence-free condition
($\nabla \cdot {\bf v} =\nabla \cdot {\bf B}= 0$).
We use up to $384^3$ collocation points.
At $t=0$, 
the magnetic field has either only uniform component (when $B_0 \not=0$) or
only random components (when $B_0 =0$) and the velocity has a support
between $2\leq k \leq 4$ in wavevector space.

Either physical viscosity (and diffusion) or hyperviscosity (and 
hyperdiffusion)
is used for dissipation terms (see Table \ref{table_2}).
The power of hyperviscosity
is set to 3 or 8, such that the dissipation term in the above equation
is replaced with
\be
 -\nu_n (\nabla^2)^n {\bf v},
\ee
where $n=$ 3 or 8 and $\nu_n$ is determined from 
the condition $\nu_n (N/2)^{2n} \Delta t
\approx 0.5$.
\footnote{When a high-order hyperdiffusion is used,
    the spectral properties near the dissipation cut-off are affected 
    by a strong bottleneck effect.
    The bottleneck effect affects high-$k$ Fourier components.
    Since we study mostly the behavior of the {\it total} $v^2$ and $b^2$ which depend mostly on 
    small-$k$ Fourier components, we believe the bottleneck effect is not a serious issue in our study.
}  
Here $\Delta t$ is the time step and $N$ is the number
of grid points in each direction.
The same expression is used for the magnetic dissipation term.
We list parameters used for the simulations in Table \ref{table_2}.
We use the notation XY-$B_0$Z, 
where X = 384, 320, 256, 144, 96, or 64 refers to the number of grid points 
in each spatial
direction; Y = H or P refers to hyper- or physical viscosity;
Z refers to the strength of the external magnetic fields.
Diagnostics of our code can be found in Cho \& Vishniac (2000a).

We use the following notations: 
\begin{description}
\item{$B=\sqrt{B_0^2+b^2}$:} total magnetic field strength or its Alfven speed. 
\item{$B_0$:} mean magnetic field or its Alfven speed.
\item{$b$, $v$ $(=v_{rms})$:} the average r.m.s. random magnetic field and velocity.
      Average is taken after
      turbulence reaches a statistically stationary state.
\item{$b^{(0)}$, $v^{(0)}$:} the zeroth-order magnetic field and 
        velocity when $B_0=0$.
\item{$E_v(k)$:} $E_v(k)=(1/2)\sum_{k-0.5 \leq k^{\prime} < k+0.5} 
| \hat{\bf v}_{{\bf k}^{\prime}} |^2$, where 
$\hat{\bf v}_{{\bf k}^{\prime}}$ is the Fourier component of velocity.
We define the magnetic energy spectrum $E_b(k)$ similarly.
\end{description}

\section{Small-scale turbulence dynamo in the very weak mean-field limit}
\subsection{Expectations}
One of the most important issues in this regime is the generation of 
small-scale random fields from the large-scale regular fields. 
Since large-scale regular magnetic fields are observed in almost all astrophysical objects, this issue is of great importance in astrophysics. 
  Cho \& Vishniac (2000a; see also Kulsrud \& Anderson 1992; Kulsrud et al.~1997) 
argued that magnetic energy in this regime
grows through field line stretching and that there are two stages of 
magnetic field amplification. 
During the first stage,
stretching is most active near, or somewhat larger than, the dissipation
 scale (spectral cut-off scale) and the magnetic energy spectrum peaks at this 
scale. 
It is clear that magnetic energy
grows exponentially during this stage and that the characteristic 
timescale is the eddy turnover time at the dissipation scale.
As the magnetic energy grows, the magnetic back reaction becomes important
 at the dissipation scale. 
When energy equipartition between kinetic energy
and magnetic energy is reached at this scale, 
the stretching rate slows down and a second stage of slower growth begins. 
Fig.~3 of Cho \& Vishniac (2000a)
shows that during this stage the peak of the magnetic power spectrum moves to larger scales.  
Using phenomenological arguments similar to the ones above, 
Schekochihin \& Cowley (2007) argued
that magnetic energy grows linearly during the second stage:
\begin{equation}
 \frac{ d B^2 }{ dt } \sim \epsilon,
\end{equation}
where $\epsilon$ is the energy injection rate, which should be equal to
the total energy dissipation rate in a statistically stationary state.
The linear stage of magnetic energy growth ends when 
stretching on the energy injection scale
becomes suppressed, which occurs when the magnetic energy density
becomes comparable to the kinetic energy density.

\subsection{Growth rate at the linear stage}
\begin{figure*}[h!t]
\includegraphics[width=.45\textwidth]{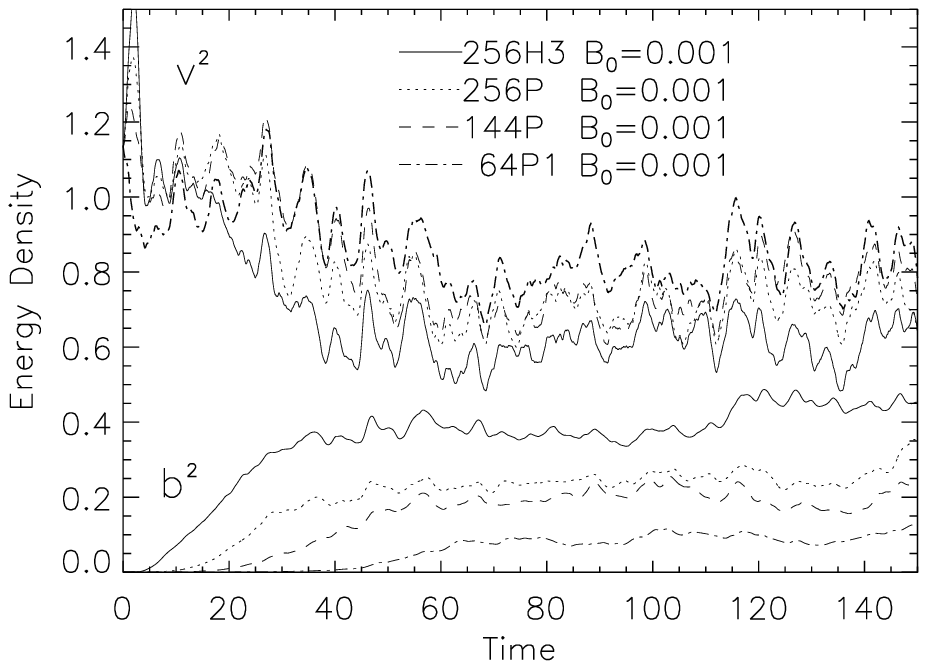}  
\includegraphics[width=.45\textwidth]{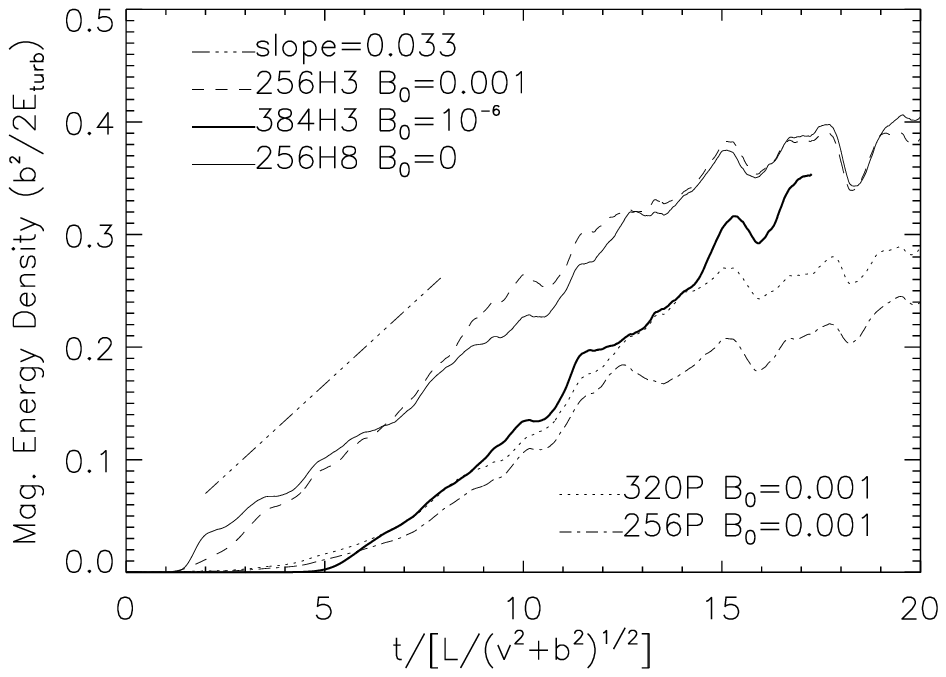} 
\caption{Time evolution of kinetic and magnetic energy densities.
  {(Left panel)}: The level of magnetic energy at the saturation level strongly
  depends on the value of magnetic diffusivity (=viscosity). 
  {(Right panel)}: Comparison of magnetic energy growth rates. 
  The growth rates at the linear growth stage are similar.
  In the case of 256H8-B$_00$, the mean field is zero and the magnetic energy spectrum
   at t=0 peaks near $k\sim 70$.
   In all other cases, only a weak mean field is present at t=0.
   Note that runs $256H8-B_00$ (i.e. run with no imposed field)
   and $256H8-B_010^{-3}$ show similar growth rates
   and also similar final saturation levels.
\label{fig:gen}}
\end{figure*}

Let us first consider the growth rate of magnetic energy during the linear stage, 
which
has an important consequence for the strength of the magnetic fields
in the large-scale structure of the universe (see Ryu et al.~2008).
Fig.~\ref{fig:gen} shows time evolution of $v^2$ and $b^2$.
All simulations started with mean magnetic field strength of
$B_0=0.001$. No random magnetic component was present at the beginning
of the simulation.
As simulations go on, random magnetic components are generated
and magnetic energy grows through stretching of magnetic field lines.
The growth of magnetic energy is slow when viscosity (and magnetic diffusivity)
is high.  For example, Run 64P1-$B_010^{-3}$ shows substantially slower
growth rate than Run 256P-$B_010^{-3}$. 
The growth rate seems to show a convergence as viscosity decreases.
For example, there is no big
difference in magnetic field growth rate between Run 256P-$B_010^{-3}$
and Run 256H3-$B_010^{-3}$.

We compare the growth rates using simulations with different parameters.
Right panel of Fig.~\ref{fig:gen} shows that the magnetic energy growth rates 
during the linear
growth stage are very similar. 
Note that we use proper normalization for both horizontal and vertical axes.
We plot only high resolution runs.
Runs with physical viscosity (and magnetic diffusion) 
show slightly smaller slopes,
which is reasonable. The strength of mean magnetic field does not seem to
affect the linear growth rate. However, we can clearly observe that,
when the mean field is weaker, the onset of the linear stage occurs later.
It is also worth noting that even the run with zero mean field ($256H8-B_00$)
shows a similar linear growth rate.
In the run with zero mean field ($256H8-B_00$), only small-scale ($k\sim 70$) 
random magnetic field is present at t=0.

In code units, the linear growth stage ends at $t\sim 40$.
The values of $B^2$ at that time is $\sim 0.4$.
Therefore, the slope during the linear growth stage 
is around 0.01.
But, when we represent the slope in terms of normalized energy 
density and time, 
we obtain different slopes:
\begin{equation}
      \frac{ B^2(t) }{ 2 E_{turb} } \sim 0.07 
    \frac{ \epsilon }{ 2E_{turb} (v/L) } \frac{ t }{ L/v }
     + \mbox{const.}
\end{equation}
or
\begin{equation}
      \frac{ B^2(t) }{ 2 E_{turb} } \sim 0.033  
    \frac{ t }{ L/\sqrt{v^2+b^2} }
     + \mbox{const.},                 \label{eq:grate}
\end{equation}
where we use $\epsilon \sim 0.16$, $L\sim 2.5$, $v\sim 0.9$, $v^2+B^2 \sim 1.0$, and 
$E_{turb}=(v^2+B^2)/2 \sim 0.5$ (see right panel of Fig.~1).\footnote{
A careful examination of Fig.~1 reveals that, although $v^2$ decreases
and $b^2$ increases during the growth stage of magnetic energy, the sum of
$v^2$ and $b^2$ does not change much. In hyper-viscosity runs,
the value of $v^2 + b^2$ is around 1 all the time.
This is why we use $\sqrt{v^2 + b^2}$ ($\approx \sqrt{v^2 + B^2}$) in
eq.~(\ref{eq:grate}).} 
A similar linear growth rate has been observed
in a recent work by Ryu et al. (2008), in which they derived strength
of magnetic fields in the large-scale structure of the universe.
        In their model, the linear growth rate derived from a simulation
        plays an essential role.

\subsection{Saturation level}

Using data with relatively low numerical resolutions, 
Cho \& Vishniac (2000a) showed that, in the limit of 
$\nu~(=\eta) \rightarrow 0$, 
the magnetic energy density in the saturation stage
is comparable to the kinetic energy density. 
This is consistent with the fact that
magnetic fields are amplified through field line stretching:
as we mentioned before, 
stretching of magnetic field
at the energy injection scale is suppressed only when 
the magnetic energy density
becomes comparable to the kinetic energy density.
In this subsection we present results with higher numerical resolutions.

We list the energy densities in the saturation stage in Table 1. 
 We obtained $v^2$, $b^2$, $\epsilon$, and $D_M$ by averaging over 
($t_1$, $t_2$). 
Here $D_M$ is the magnetic energy dissipation rate.
It is important to note that
 these time averages are taken after the turbulence has reached a 
stationary state. Conclusions based on these averaged values do not apply to 
the initial 
growth phase of the magnetic field. 

\begin{figure*}[h!t]
\includegraphics[width=.45\textwidth]{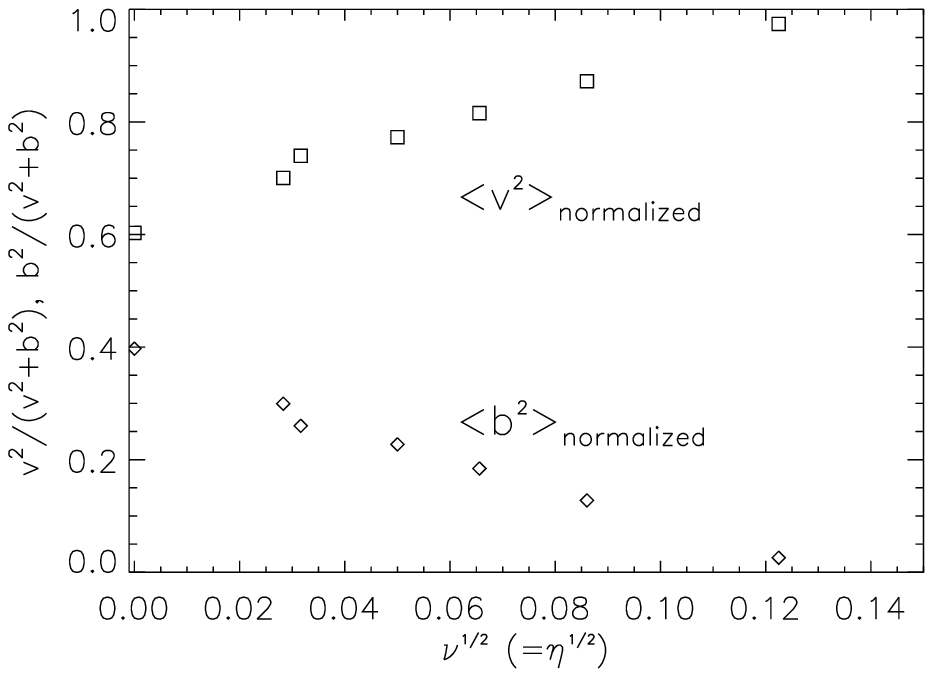}   
\caption{Normalized average kinetic and magnetic energy densities.
         Average is taken after turbulence has reached a saturation state.
\label{fig:gen2}}
\end{figure*}

Fig.~\ref{fig:gen2} shows $v^2$ and $b^2$ as 
functions of $\nu ~(=\eta)$.\footnote{
     The use of $\sqrt{\nu} ~(=\sqrt{\eta})$ for horizontal axis is not motivated by
     theoretical considerations, but by clarity of presentations.
}
All the simulations shown in left panel of Fig.~1
have similar kinetic energy densities. 
 However, the magnetic energy density obviously depends on the ohmic diffusivity $\eta$.
When $\nu (=\eta)$ goes to zero, the magnetic energy seems to 
approach to $\sim$40\% of the total energy. This is somewhat larger than
the value obtained by Haugen et al.~(2003), where $\sim 30\%$ of the total
energy is magnetic.
The discrepancy may stem from the fact that we use hyper-viscosity
to achieve a very small viscosity (and magnetic diffusion).

\begin{figure*}[h!t]
\includegraphics[width=.45\textwidth]{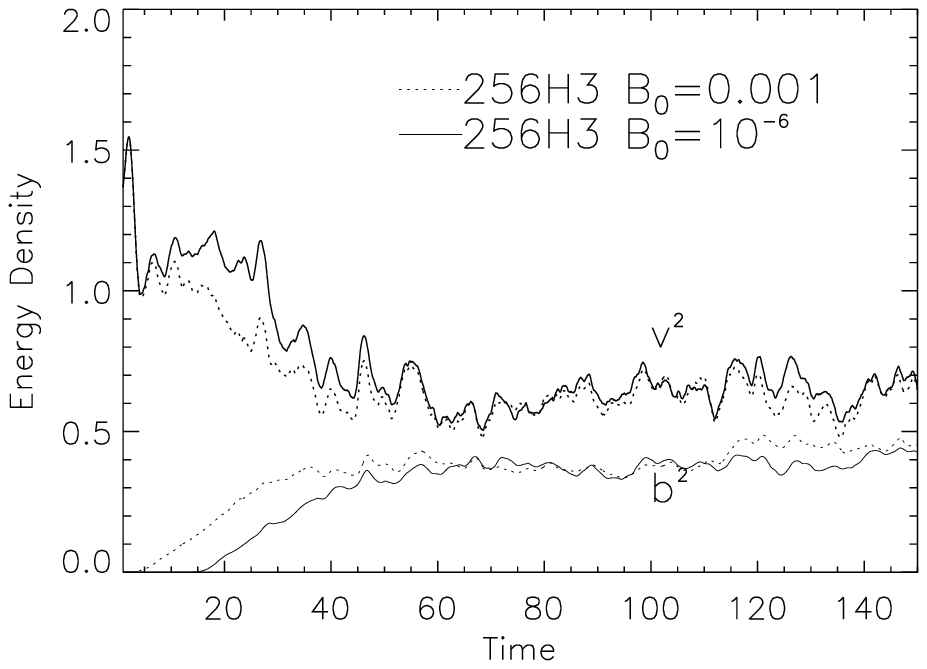}   
\includegraphics[width=.45\textwidth]{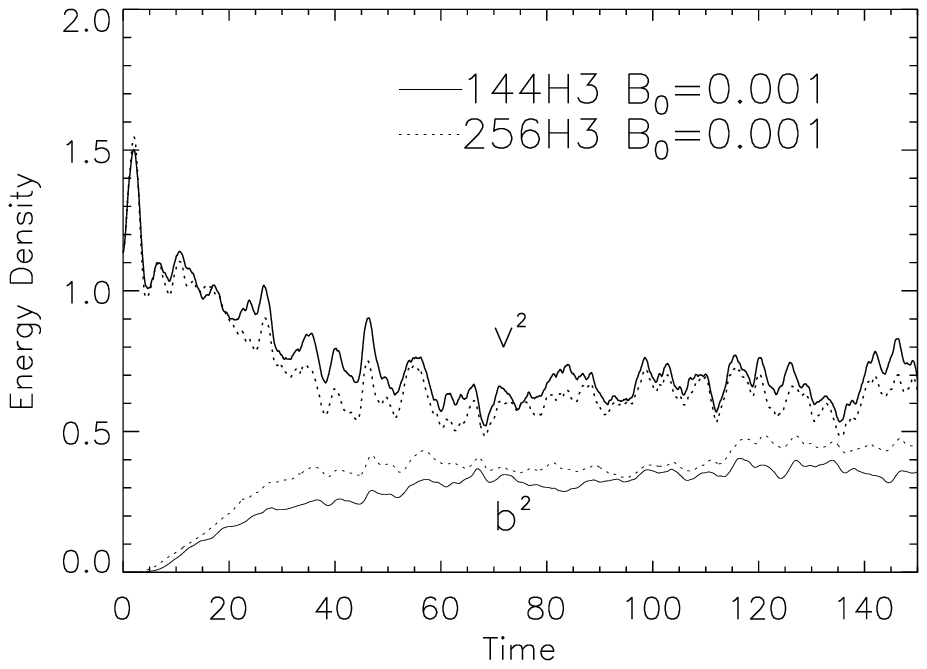}  
\caption{Comparison of runs.
         {\it (Left panel):} These runs with hyperdiffusion shows that
         initial strength of the mean magnetic field does not
         strongly affect the growth rate and
         saturation level of magnetic field. 
         However, amplification of magnetic field (in fact, onset of
         linear growth stage) is delayed when the
         seed mean magnetic field is weak.
         The delay will be negligible when 
         the dissipation scale is very small compared with the
         energy injection scale (see \S\ref{sect_actualastro} for details).
         {\it (Right panel):} These two runs with hyperdiffusion 
         show  that numerical resolution also does not strongly
         affect the growth timescale, although two runs show
         slightly different levels of energy saturation.
\label{fig:tests}
}
\end{figure*}

Runs shown in left panel of Fig.~\ref{fig:tests} show that the order of 
hyper-diffusion does not strongly
 affect the growth rate and saturation level of
magnetic energy.
Runs shown in right panel of Fig.~\ref{fig:tests} show that numerical resolution
slightly affects the saturation level of magnetic energy.


\subsection{Exponential growth stage}  \label{sect:growthB}
\begin{figure*}[h!t]
\includegraphics[width=.45\textwidth]{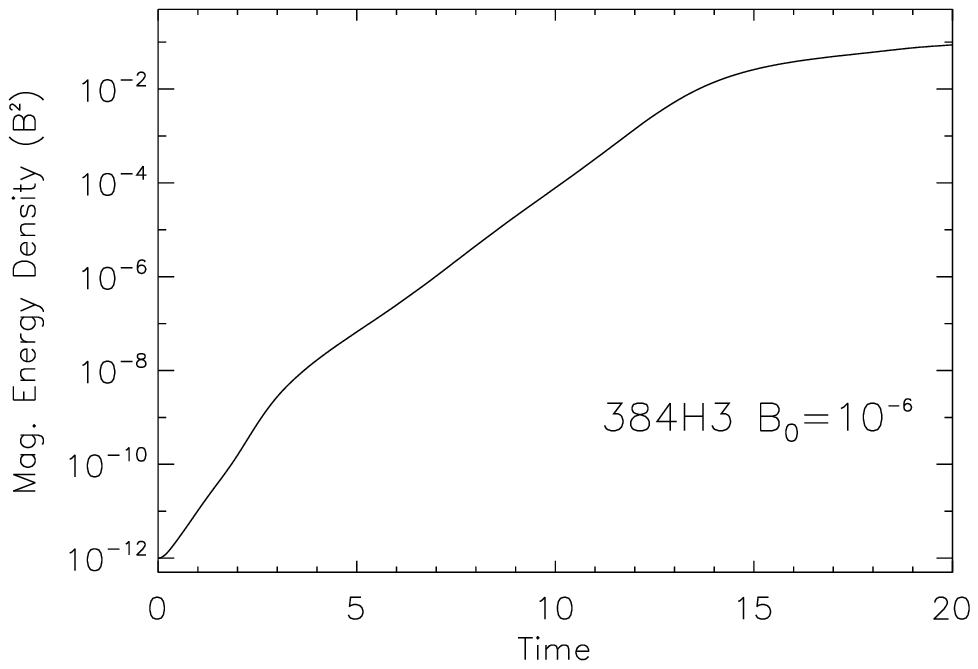}   
\includegraphics[width=.45\textwidth]{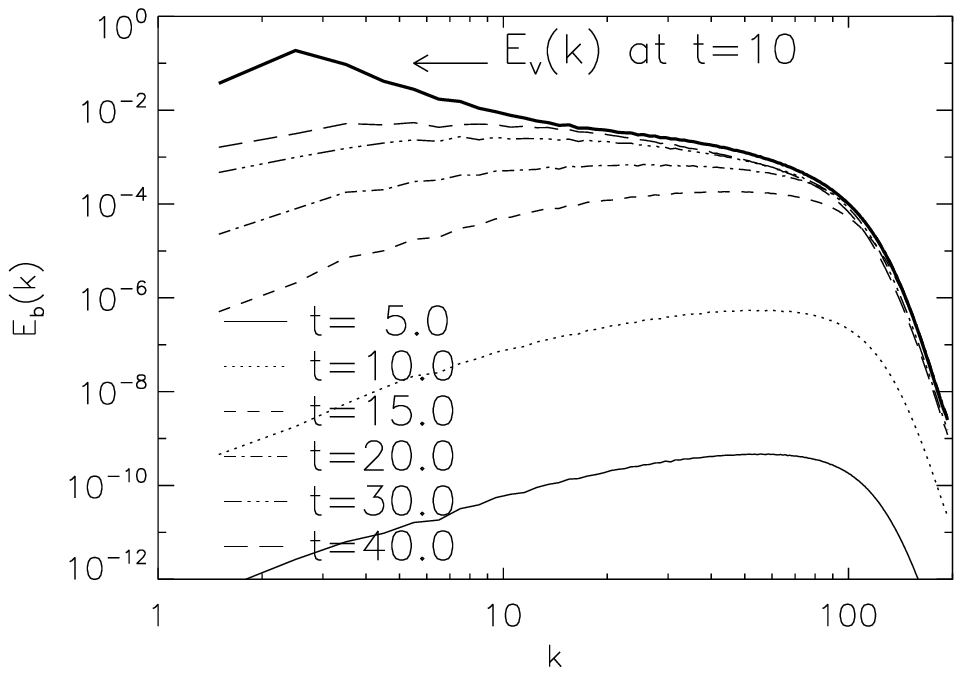} 
\caption{Time evolution of Run 384H3-B$_010^{-6}$.
         This run with hyperdiffusion shows rapid exponential growth
         at the beginning, which is  
         due to the stretching of magnetic field lines near the
         dissipation cut-off at $k_{d} \sim 100$.
         During this phase, magnetic spectra peaks at $k\sim 80$ and
         moves upwards.
         After $t\sim 15$, magnetic energy
         becomes comparable to the kinetic energy 
                     at $k_{d} \sim 100$ and
         the initial rapid exponential growth phase ends.
\label{fig:10_12}
}
\end{figure*}

Magnetic field is amplified through 
field line stretching, which  
is initially most active near the dissipation
 scale.
As a result, magnetic energy will grow exponentially at the beginning.
To see this stage more clearly, we plot time evolution of magnetic energy
in logarithmic scales.
We use Run 384H3-B$_010^{-6}$.
As expected, Fig.~\ref{fig:10_12} clearly shows this exponential growth stage.
The strength of the mean field is $10^{-6}$ and the dissipation scale is
around $k\sim 100$.
When $t < 15$, the growth rate is exponential and magnetic energy spectrum peaks
near $k\sim 100$. 
At $t\sim 15$, energy equipartition is reached at $k \sim 100$ and 
the exponential growth stage ends.
After $t\sim 15$, 
the linear growth stage begins and the peak of magnetic energy spectrum
moves to smaller $k$'s.

\section{Effects of the mean field  (B$_0$)}
In this section, we consider a similar numerical set-up as in the previous section: 
only a mean magnetic field $B_0$ is present at t=0.
Then,  
turbulent motions generate fluctuating magnetic field ${\bf b}$
and, as time goes on, the strength of the fluctuating field grows.
After a certain amount of time, which depends on $\eta$ and $B_0$,
turbulence reaches a statistically stationary state.
In this section, we study how the strength of the mean field ($B_0$)
affects energy densities and other properties of turbulence.
As we will see in this section,
when the mean field is stronger, magnetic energy rises faster,
the system reaches 
the statistically stationary
state more quickly, and
magnetic energy density at the saturation stage
is higher,
while kinetic energy density is smaller.
This result is consistent with an earlier result by
Tao, Cattaneo, \& Vainshtein (1993).
Although their main conclusion is the suppression of $\alpha$ effect in the
presence of mean fields, their figures show that
magnetic energy density at the saturation stage is higher when the mean
field is stronger.
Our result is also consistent with those by Schekochihin et al. (2007),
where they studied the effects of mean fields for different magnetic Prandtl
numbers.

\subsection{Scaling of energy densities, $v^2$ and $b^2$}

\begin{figure*}[h!t]
\includegraphics[width=.45\textwidth]{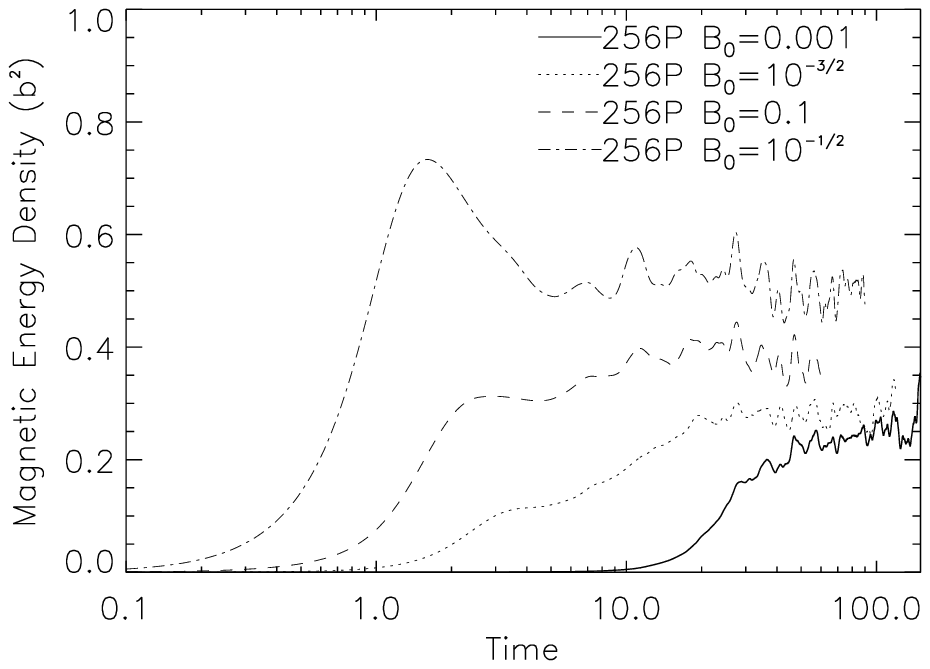}   
\includegraphics[width=.45\textwidth]{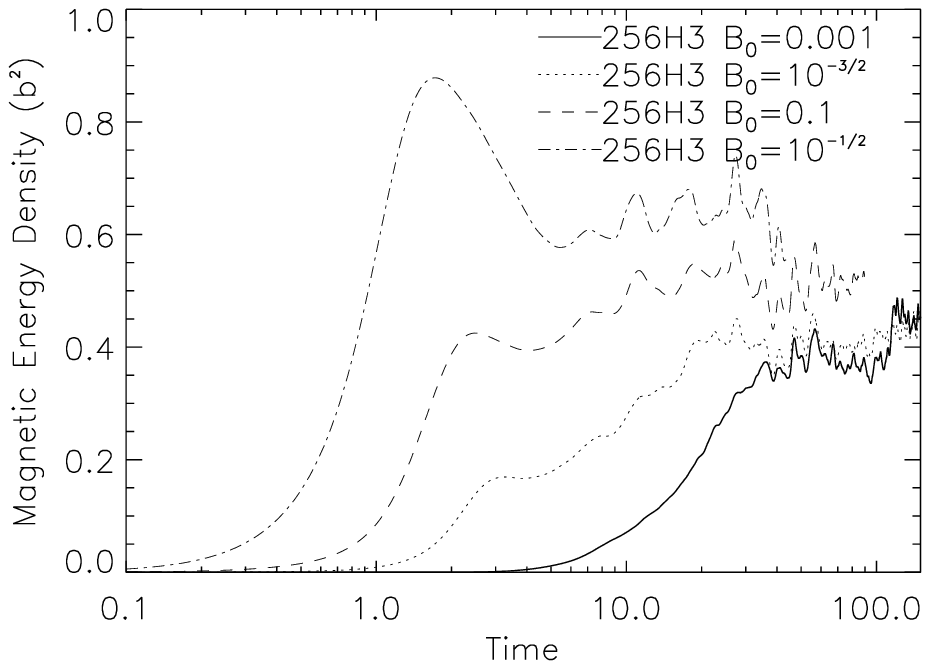}  
\caption{
Time evolution of magnetic energy density. 
The stronger the mean magnetic field, the higher the magnetic energy
 at the saturation stage. 
The growth time is shorter when the mean magnetic field is stronger.
{\it (Left panel):} Runs with physical diffusion.
{\it (Right panel):} Runs with hyper-diffusion.
\label{fig:effB}
 }
\end{figure*}

\begin{figure*}
\includegraphics[width=.45\textwidth]{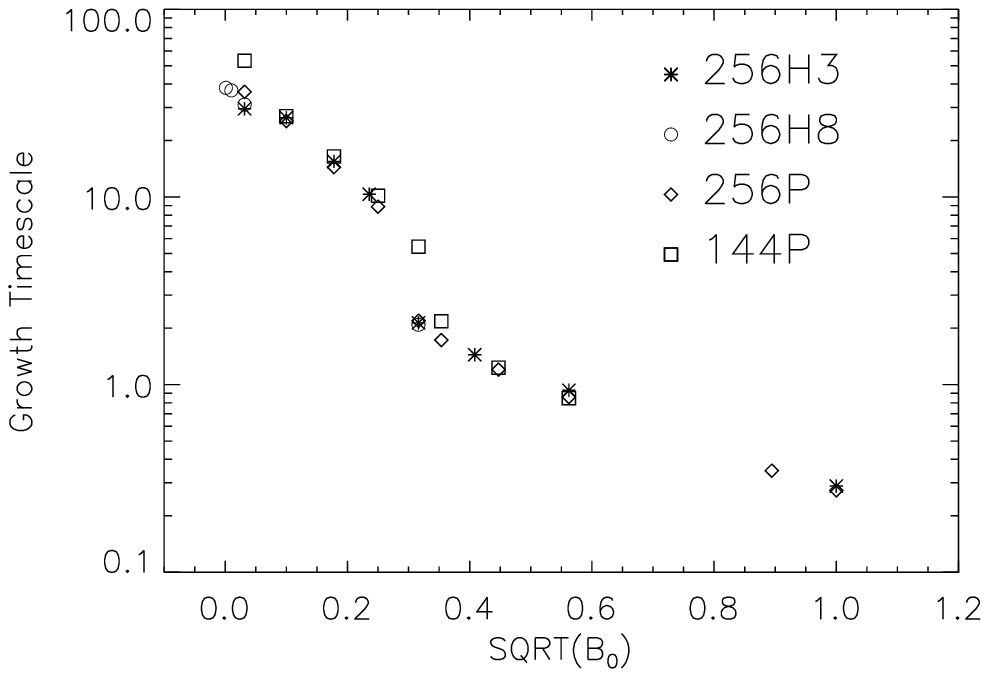}  
\caption{
Growth timescale. The y-axis is the time in code units when the 
{\it fluctuating} magnetic energy first reaches 80\% of the
average value of the fluctuating magnetic energy at the saturation stage.
\label{fig:growthtime}
}
\end{figure*}

Fig.~\ref{fig:effB} shows time evolution of fluctuating
 magnetic energy density.
When the mean field gets stronger, the saturation level increases
and the growth time becomes shorter.
Runs with both physical and hyper-viscosity show similar behavior.
In Fig.~\ref{fig:growthtime}, we explicitly measure the growth timescale,
which we define the time the energy of the random magnetic field reaches 
80\% of its saturation value.
Roughly speaking, Fig.~\ref{fig:growthtime} implies that
the growth time scales with $\sqrt{B_0}$.
Numerical resolution or the form of viscosity is not important unless
the mean field is extremely small.

\begin{figure*}
\includegraphics[width=.45\textwidth]{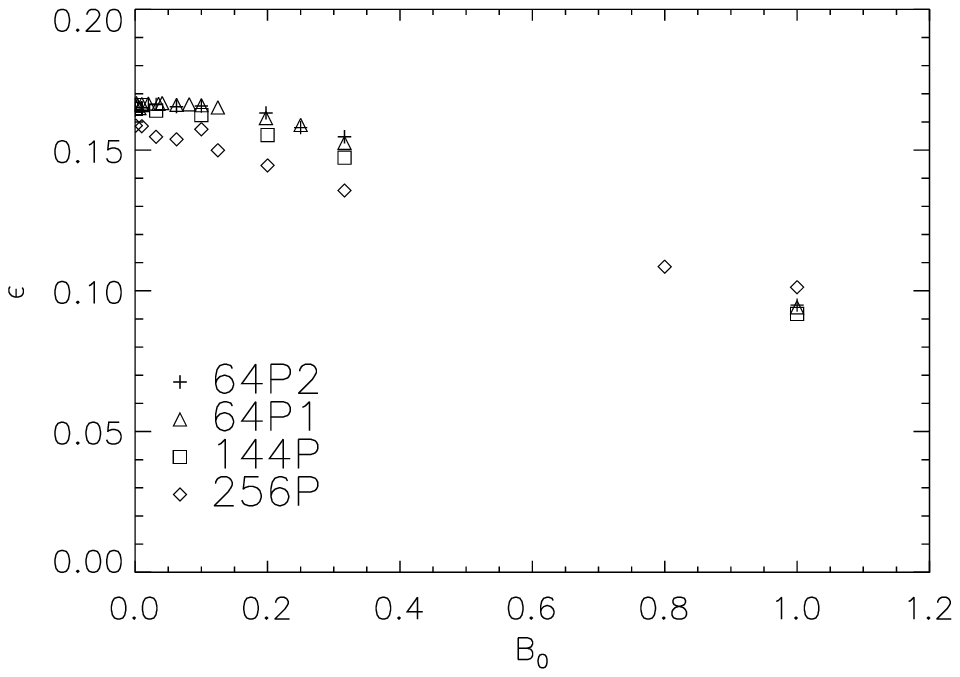}  
\caption{
Average values of the total energy dissipation 
rate.
\label{fig:eps}
}
\end{figure*}

As we can see in Fig.~\ref{fig:eps},
all our simulations, except runs with $B_0 \sim 1$, have similar energy dissipation/injection rates $\epsilon$ 
($= {\bf f}\cdot{\bf v}$).
when the mean magnetic field is very weak, the rate is $\sim 0.16$ and
only weakly depends on $\nu ~(=\eta)$.
The rate declines slowly as the external field gets stronger.
We list the total energy dissipation rate and the magnetic energy
dissipation rate in Table 1.
In 256P runs, the ratio $D_M/\epsilon$ is $\sim 0.7$ in weak-mean field cases
and it drops to $\sim 0.55$ in strong mean field cases (e.g.~256P-$B_00.8$
and 256P-$B_01$).
The result for weak mean-field cases is consistent with Haugen, Brandenburg,
\& Dobler (2004).

\begin{figure*}[h!t]
\includegraphics[width=.45\textwidth]{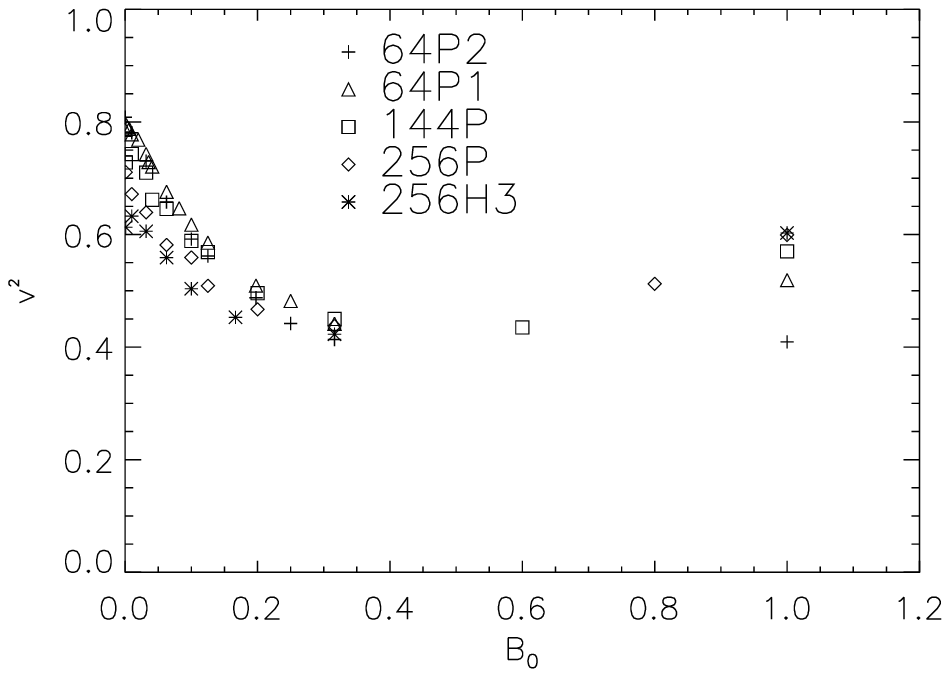}  
\includegraphics[width=.45\textwidth]{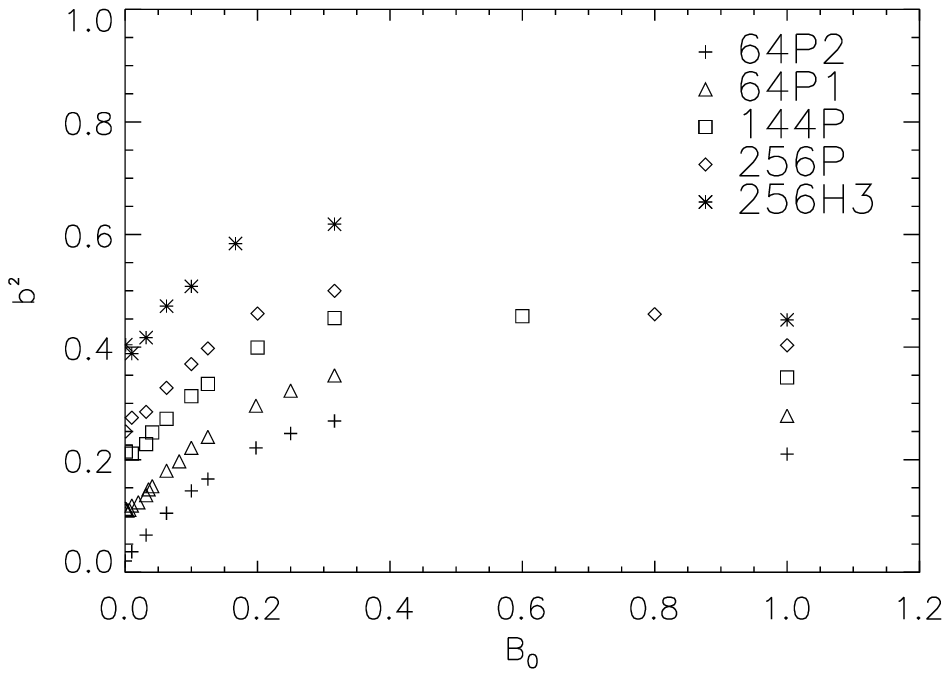}  
\caption{Average values of $v^2$ and $b^2$.
Note that $b^2$ is the energy of fluctuating magnetic field.
When $B_0 > 0.1$, magnetic energy, ($B_0^2+b^2$)/2,
can be larger than kinetic energy, $v^2$/2, for 256H3 runs.
\label{fig:v2b2}
}
\end{figure*}

We measure average energy densities at the statistically stationary state
and list them in Table 1.
In Fig.~\ref{fig:v2b2}, 
we plot the fluctuating energy densities, $v^2$ and $b^2$ 
($=B^2-B_0^2$), as functions of mean field strength $B_0$.

First, from the Figure (and Table 1), 
we note that there is no difference between turbulence with
no external magnetic field and one with very weak external fields.
(For example, compare 256H8-B$_00$ and 256H8-B$_010^{-6}$, and
              96P-$B_00$ and 96P-$B_010^{-3.5}$.)
When the external fields are very weak, the kinetic and
magnetic energy densities go smoothly to the zero external field limit. 
{}From the Figure, we also note that the kinetic energy densities are not
very sensitive to the value of $\nu~(=\eta)$ while the magnetic
energy densities do show a strong dependence on $\nu~(=\eta)$
(see Cho \& Vishniac 2000a for more discussion for $B_0 \approx 0$ limit).
In the case of hyperviscosity simulations, the ratio $b^2/v^2$ 
for $B_0 \rightarrow 0$
is greater than 0.6.
When $B_0 > 0.1$, magnetic energy, ($B_0^2+b^2$)/2,
can be larger than kinetic energy, $v^2$/2, for 256H3 runs.

Second, when the external fields are not very strong 
(say, $B_0 \lesssim 0.2$),
the fluctuating energy densities follow 
\begin{eqnarray}
  b^2 \propto b^{(0)2} + c_b vB_0 , \label{bsq} \\
  v^2 \propto v^{(0)2} - c_v vB_0, \label{vsq}
\end{eqnarray}
where $c_b$ is almost independent of $\nu ~(=\eta)$ and $c_v$ 
weakly depends on $\nu ~(=\eta)$
and superscript `(0)' denotes values for $B_0=0$.
Equation (\ref{bsq})
implies that 
\be
  b > (vB_0)^{1/2} \gg B_0.
\ee


\subsection{Scaling of total energy}

\begin{figure*}[h!t]
\includegraphics[width=.45\textwidth]{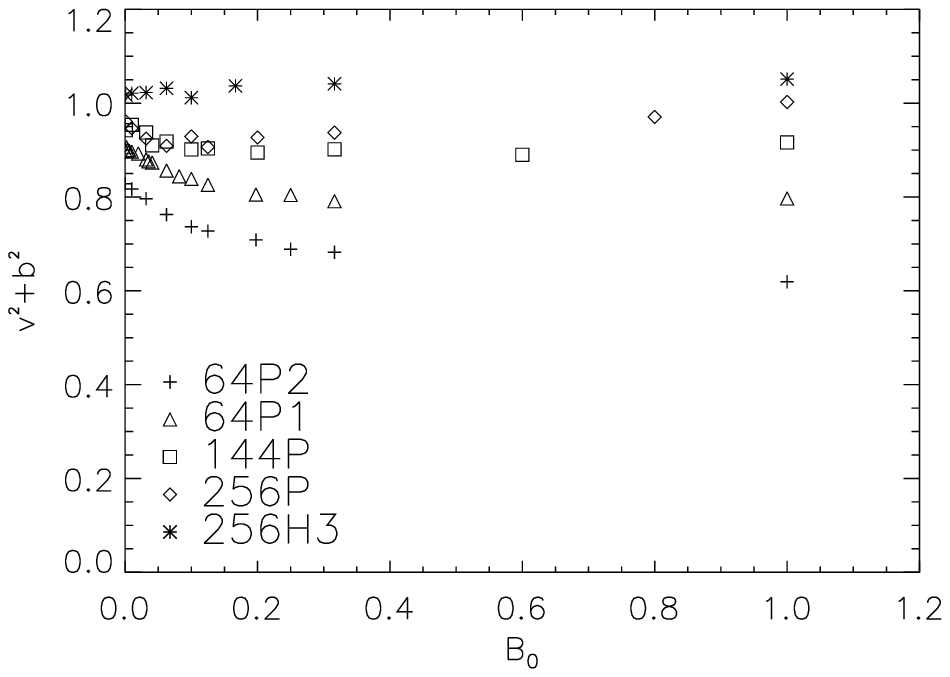}   
\caption{Total energy density and $B_0$.
Total energy does not show a strong dependence
on the mean field strength.
\label{fig:vplusb}
}
\end{figure*}

One consequence of the above scaling relations 
is that the sum $v^2 + (c_v/c_b) b^2$ is approximately
independent of $B_0$ when
the external fields are not very strong:
\be
  v^2 + (c_v/c_b) b^2 \approx v^{(0)2}+ (c_v/c_b) b^{(0)2}. \label{etot}
\ee
Fig.~\ref{fig:vplusb} shows that total energy ($v^2+b^2$) does
not strongly depend on $B_0$. 
The constancy of the total energy is especially good for
runs with high numerical resolutions (256P and 256H3 runs).
Therefore the Figure implies that $(c_v/c_b)\approx 1$ for 256P and 256H3 runs.
However, the value of  $(c_v/c_b)$ is less than 1 for lower resolution runs.
Since $v^2 + b^2$ is virtually independent of $B_0$, we can use
the quantity for normalization. For example, we may define
generalized large-scale eddy turnover time as $L/(v^2+b^2)^{1/2}$.

\subsection{Scaling of energy transfer rate}

\begin{figure*}
\includegraphics[width=.45\textwidth]{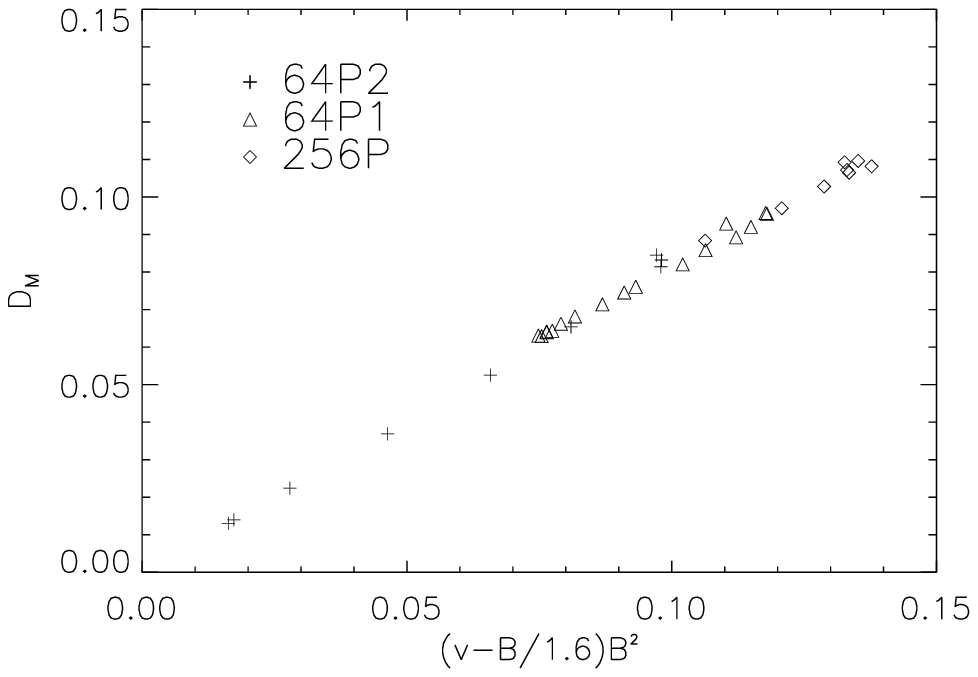}  
\caption{
Stretching effect. Magnetic fields are amplified through field line
   stretching. The stretching rate is proportional to $vB^2$, which 
   may be the stretching rate for passive vector fields, minus
   $BB^2$, which may be regarded as magnetic back reaction, times a constant.
   The value of $B_0$ ranges from 0 to $10^{-0.5}$. We use runs with
   physical viscosity only.
   Runs with $B_0 \sim 1$, do not follow this relation.
\label{fig:stretching}
}
\end{figure*}

Cho and Vishniac (2000a) showed that, 
when the external fields are weak/zero, 
magnetic fields are amplified through field line stretching:
\be
  D_M \propto (v-cB)B^2/L, \label{stretching}
\ee
where $D_M$ is magnetic dissipation and $c\approx 1/1.6$.
Fig.~\ref{fig:stretching} confirms that this result is 
also true even when the external
fields are strong. 
In the stationary state,
the magnetic dissipation ($D_M$) is balanced by the net energy
transferred to the magnetic field from the velocity field.
The right-hand side therefore tells us that the net energy
transferred to the magnetic field is proportional to large-scale eddy
turnover rate ($v/L$) minus an 
Alfv\'{e}nic frequency ($B/L$) times a constant.
The large-scale eddy turnover rate is equal to the stretching rate
of the magnetic field when the back reaction is zero.
We identify the second term on the right-hand side of this 
equation as the effect of the magnetic back reaction.

\subsection{Energy spectra}

\begin{figure*}
\includegraphics[width=.45\textwidth]{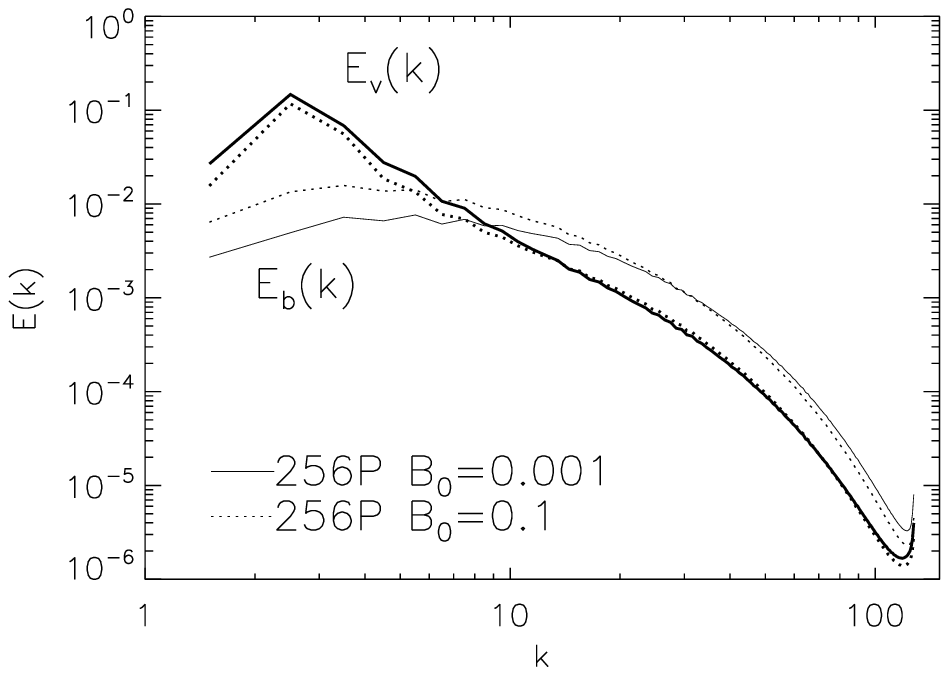}  
\caption{
 Energy spectra. Kinetic energy spectrum
peaks at the driving scale (=energy injection scale).
When the mean field gets stronger
 kinetic spectrum near the driving scale goes down.
The magnetic energy spectrum peaks at wavelengths larger than the
energy injection scale. The location of the peak moves toward smaller
wavenumber as the mean field gets stronger. 
Note that, when $B_0 < 1$, the kinetic spectrum does not show 
a well defined power law and the average slope should be steeper than
the Kolmogorov one near the energy injection scale. 
\label{fig:sp}
}
\end{figure*}

We plot energy spectra in Fig.~\ref{fig:sp}.
Kinetic spectra peaks at $k\sim 2.5$, which is independent of $B_0$.
Kinetic spectrum for the run with $B_0 = 0.1$ 
(thick solid line) has less energy at
small k's than that with $B_0 = 0.001$ (thick dashed line).
On the other hand, magnetic 
spectrum for $B_0 = 0.1$ (solid line) has more energy at
small k's than that for $B_0 = 0.001$ (dashed line).
As the mean field strength grows, 
the peak of the magnetic spectra moves from $\sim k_L/2.5$ 
($B_0\approx 0$ case) to
 $\sim k_L$ ($B_0=v_{rms}$ case), where $k_L$ is the wavenumber of
the peak of
kinetic energy spectra. 
Kinetic spectra are steeper than Kolmogorov spectrum for small $k$'s,
while magnetic spectra are flatter.
Therefore, one should be careful when using Kolmogorov spectrum 
     for MHD turbulence: the kinetic and magnetic spectra are not 
     Kolmogorov for small $k$'s when $B_0$ is weak. 
Since the slope of the total energy spectrum roughly follows 
Kolmogorov slope (see Fig.~\ref{fig:spect_hdmhd} and \S\ref{sect:hdmhd}),
the slope of kinetic spectra at small k's is a function of $B_0$:
when $B_0$ is small, kinetic spectra should be steeper than $k^{-5/3}$ and
those for larger $B_0$ are closer to $k^{-5/3}$.
When $B_0 \sim v_{rms}$, the kinetic energy spectrum
is very close to the  Kolmogorov one consistent with the 
predictions in Goldreich \& Sridhar (1995) as well as with some of the 
earlier simulations (Cho \& Vishniac 2000b; Cho et al. 2002). However, we cannot
make a stronger conclusion with the present resolution (cf. Maron \& Goldreich 2001;
Beresnyak \& Lazarian 2006; Mason et al. 2006). 


\subsection{Anisotropy}

When $B_0 \rightarrow v$, MHD turbulence tends to be anisotropic
(Shebalin et al 1983; Goldreich \& Sridhar 1995).
Here we focus on {\it global} anisotropy of MHD turbulence with respect to
the {\it mean} magnetic field ${\bf B}_0$.\footnote{
     In this subsection, we consider global anisotropy, not local anisotropy.
     That is, we calculate anisotropy with respect to the direction of
     the imposed field.
     When we calculate anisotropy with respect to directions 
     that follow the local mean field
     in smaller sub-volumes, we do obtain anisotropy for all values of $B_0$.
     This kind of anisotropy is {\it local} anisotropy.
     Scale-dependent anisotropy discussed in
     Goldreich \& Sridhar (1995) is 
     local anisotropy 
     (see Cho \& Vishniac 2000b and Cho et al. 2002 for details).
     Local anisotropy is important for 
        local physical processes (e.g.~pitch-angle scattering).
       But, what matters for external observers
        is mostly global anisotropy (or global isotropy).
 }
Let $k_{\|}$ be the wavenumber parallel to the mean field and
$k_{\perp}$ the wavenumber perpendicular to the mean field.
Matthaeus et al. (1998) showed that
the anisotropy of MHD turbulence scales linearly with
the ratio of perturbed and total magnetic strength $b/B$.
Their result is mainly for $B_0 \gtrsim v$.
No systematic study is available for $B_0 < v$ cases.

We study anisotropy of MHD turbulence in the $B_0 < v$ limit by
comparing average $k_{\|}$ and $k_{\perp}$:
\begin{equation}
 \frac{ \langle k_{\|}\rangle }{ \langle k_{\perp}\rangle }
   = 
 \frac{ \left< \int_0^{k_{max}} d^3k ~k_{\|}  
               | \hat{\bf v}_{\bf k} |^2 \right>_t }
      { \left<\int_0^{k_{max}} d^3k ~k_{\perp}
               | \hat{\bf v}_{\bf k} |^2 \right>_t },
\end{equation}
where $k_{max}$ is the maximum wavenumber and $\hat{\bf v}_{\bf k}$ is
the Fourier component of velocity at ${\bf k}$.
The result in Figure \ref{fig:ani} shows that
turbulence is virtually isotropic when $B_0 \lesssim 0.2 v$.
We note that there
is no apparent anisotropy of order $B_0/v$.
When $B_0$ becomes stronger, turbulence shows departure from isotropy.
Note again that we focus on the cases of $B_0 \leq v$ in this paper.
It is also noteworthy that our result is valid when driving is isotropic.

\subsection{Kinetic helicity}
In Fig.~\ref{fig:hk}, we plot {\it normalized} kinetic helicity,
$H_k/v^2$,
as a function of $B_0$.
The kinetic helicity, $H_k=\langle {\bf v}\cdot \nabla\times {\bf v} \rangle$,
does show suppression as $B_0$ increases.
But, constancy of $H_k/v^2$ implies that the reduction in $H_k$
is due to the reduction in $v^2$.
Therefore, 
the figure clearly shows that kinetic helicity
is not suppressed strongly.
Since we can write $H_k=\langle{\bf v}\cdot \nabla \times {\bf v}\rangle
 \sim k_{peak,H_k} v^2$, where $k_{peak,H_k}$ is the wavenumber at which 
kinetic helicity spectrum peaks,
the constancy of $H_k/v^2$ means the location of the peak wavenumber 
is not a function of $B_0$.
Indeed
the kinetic helicity spectra peak at the energy injection scale.

\subsection{Decay timescale}
There have been a lot of studies on the decay law of MHD turbulence.
Decay law and decay timescale of MHD turbulence are of great importance for
dynamics of molecular clouds and star formation.

Biskamp \& M\"uller (2000) discussed decay of incompressible MHD turbulence 
in the limit of zero external field cases.
However, there has been no systematic studies on the decay timescale for
various values of the mean field strengths.
Decay timescale of turbulence energy can be estimated by
\begin{equation}
  (v^2+b^2)/\epsilon.
\end{equation}
Fig.~\ref{fig:decay} shows that the rate is not very sensitive to the
strength of the mean field.
Note that, even the case the Alfven speed of the mean field ($B_0$)
 is comparable to
the the rms velocity ($v \sim 1$), the decay rate is
not much reduced.
This fact is consistent with earlier simulations of the decaying compressible MHD
turbulence (Stone et al.~1998; Mac Low et al.~1998) as well as those of incompressible one
(Maron \& Goldreich 2001; Cho et al. 2002).
We present a dimensionless dissipation coefficient
\begin{equation}
   D \equiv (\epsilon/2E_{turb})(L^{\prime}/u^{\prime}),  
\label{eq:diss}
\end{equation}
where $\epsilon$ is the energy dissipation rate, 
$E_{turb}=(v^2+b^2)/2$, $3{u^{\prime}}^2 = v^2$, and
\begin{equation}
  L^{\prime}=\frac{\pi}{ 2 {u^{\prime}}^2 } \int k^{-1}E_{v}(k) dk.
\end{equation}
For 256P MHD runs the value of $L^{\prime}$ is around $\sim 0.76$ 
and weakly depends
on $B_0$. For example, 
$L^{\prime}=$ 0.76 (256P-$B_010^{-3}$ and 256P-$B_010^{-1.5}$),
0.72 (256P-$B_010^{-1}$), 0.68 (256P-$B_010^{-0.5}$), and 0.71 (256P-$B_01$).
The scale $L^{\prime}$ is a mathematical representation
 of the energy injection scale and the quantity $L^{\prime}/u^{\prime}$
is a kind of large scale eddy turnover time.
Since $L^{\prime} \sim 0.76$, $u^{\prime}\sim v/\sqrt{3}$, 
for 256P MHD runs, we have $L^{\prime}/u^{\prime}\sim 1.5$, which
is smaller than $L/v$.
In 256P hydrodynamic run, $L^{\prime}\sim 0.62$.
Note that the definition of the dimensionless dissipation rate
 is not exactly the same as that in Kaneda et al. (2003),
in which they used
\begin{equation}
  (\epsilon/{u^{\prime}}^2)(L^{\prime}/u^{\prime}).  \label{eq:kaneda}
\end{equation}
The dimensionless dissipation coefficient is known to be around $0.5-0.6$
both for hydrodynamic turbulence (Kaneda et al. 2003) and 
MHD turbulence (Stone et al.~1998;
Haugen, Brandenburg, \& Dobler 2004; see McKee \& Ostriker 2007).
Our results are consistent with earlier findings:
the value is around $0.5-0.6$ when the mean field is weak.
This value somewhat drops when mean field is stronger.
This has to do with the fact that the Kolmogorov constant depends
on the value of the mean field (see next section).

\begin{figure*}
\includegraphics[width=.45\textwidth]{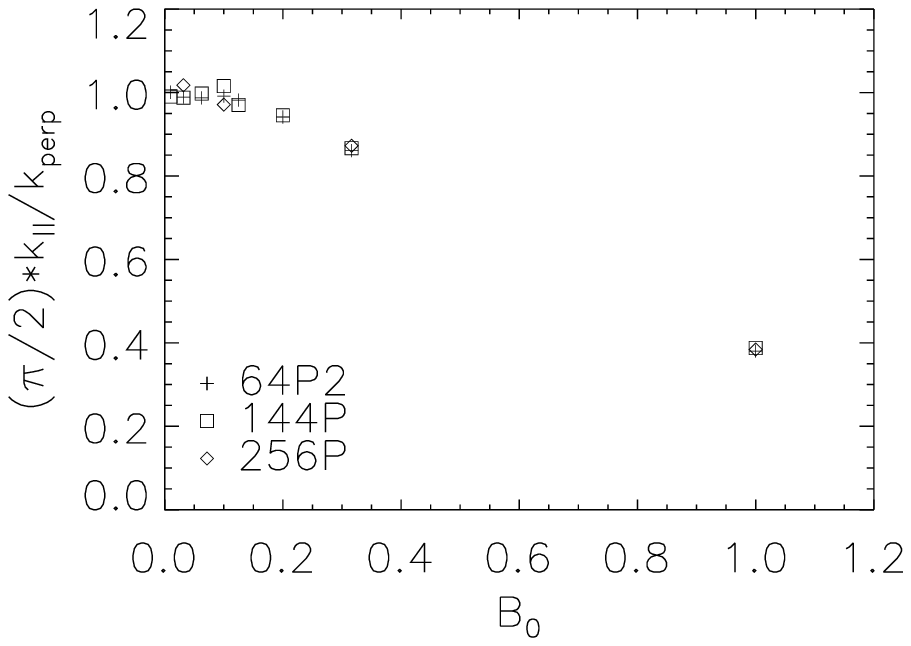}  
\caption{
Global anisotropy. Here $k_{\|}$ and $k_{\perp}$
are 
 the average wave-number
   along and perpendicular to the mean field direction, respectively.
   Turbulence remains almost isotropic when $B_0 \lesssim 0.2$.
\label{fig:ani}
}
\end{figure*}
\begin{figure*}
\includegraphics[width=.45\textwidth]{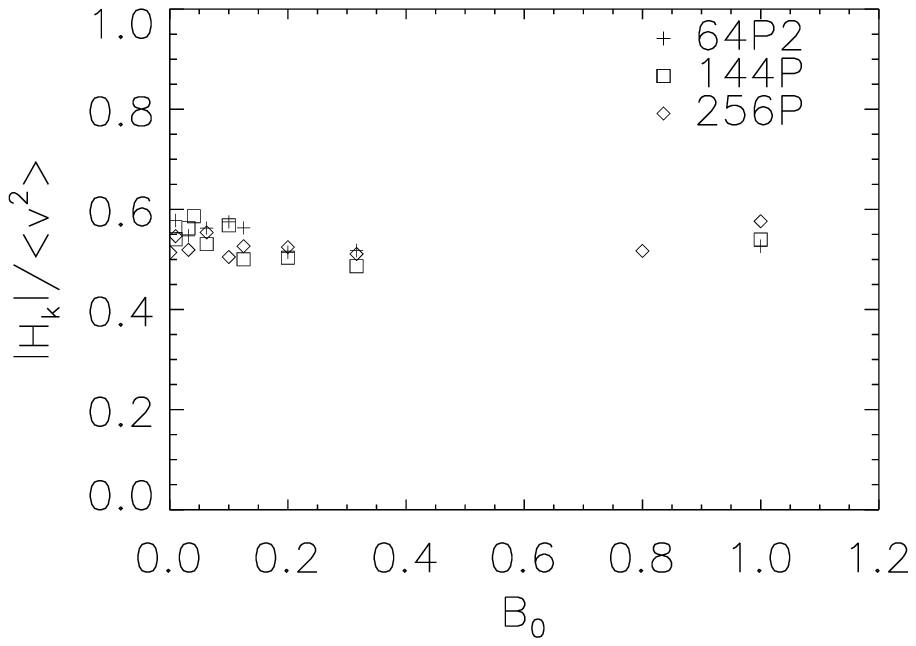}   
\caption{
   Normalized kinetic helicity $H_{K}$. 
   This Figure shows that kinetic helicity is not strongly suppressed.
\label{fig:hk}
}
\end{figure*}

\begin{figure*}[h!t]
\includegraphics[width=.45\textwidth]{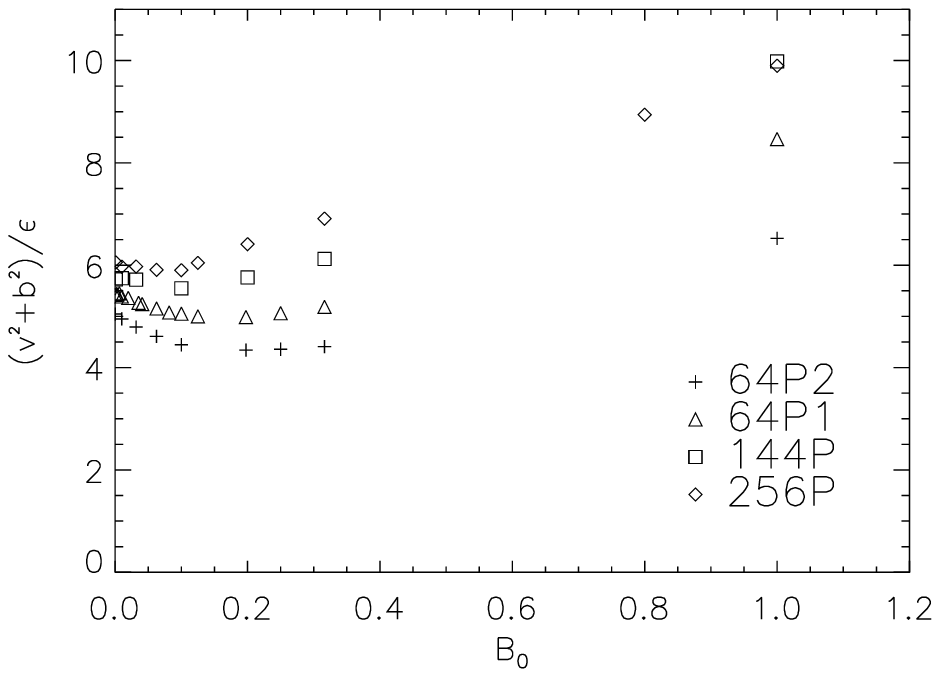}  
\includegraphics[width=.45\textwidth]{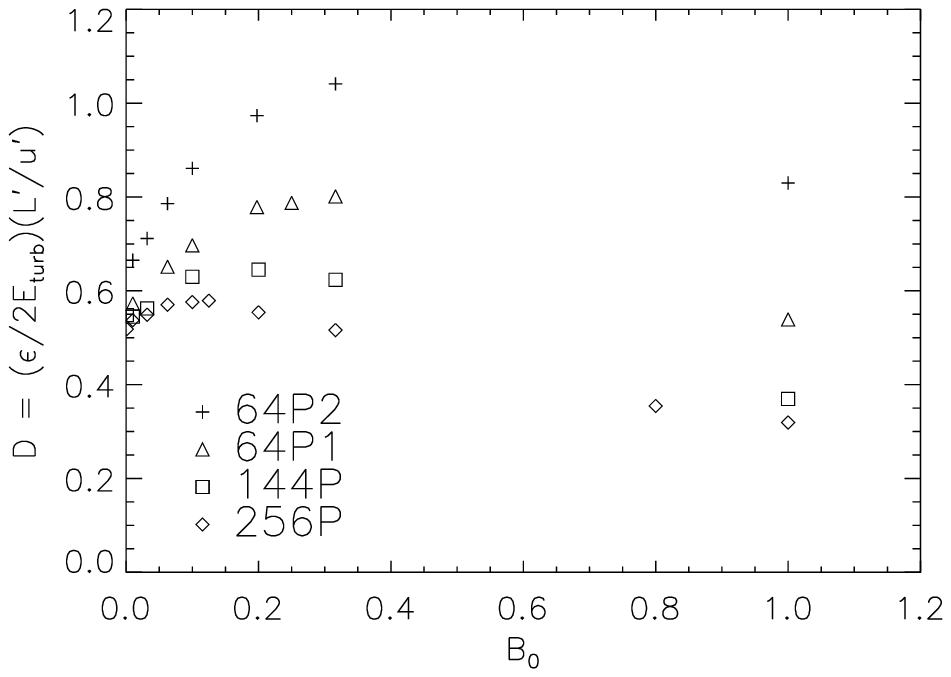} 
\caption{{\it (Left panel):} Decay timescale. 
{\it (Left panel):} Dimensionless dissipation rate. 
Note that the inverse of the dimensionless dissipation rate, $1/D$,
is a kind of normalized turbulence decay time.
\label{fig:decay}
}
\end{figure*}

\section{Comparison with Hydrodynamic simulations} \label{sect:hdmhd}
It is often claimed that MHD turbulence with weak imposed magnetic field is
similar to hydrodynamic turbulence. Is this true?
When we introduce a mean field in a fully turbulent medium,
what will happen to energy densities?
In this section, we compare hydrodynamic and MHD turbulence.

\begin{figure*}[h!t]
\includegraphics[width=.45\textwidth]{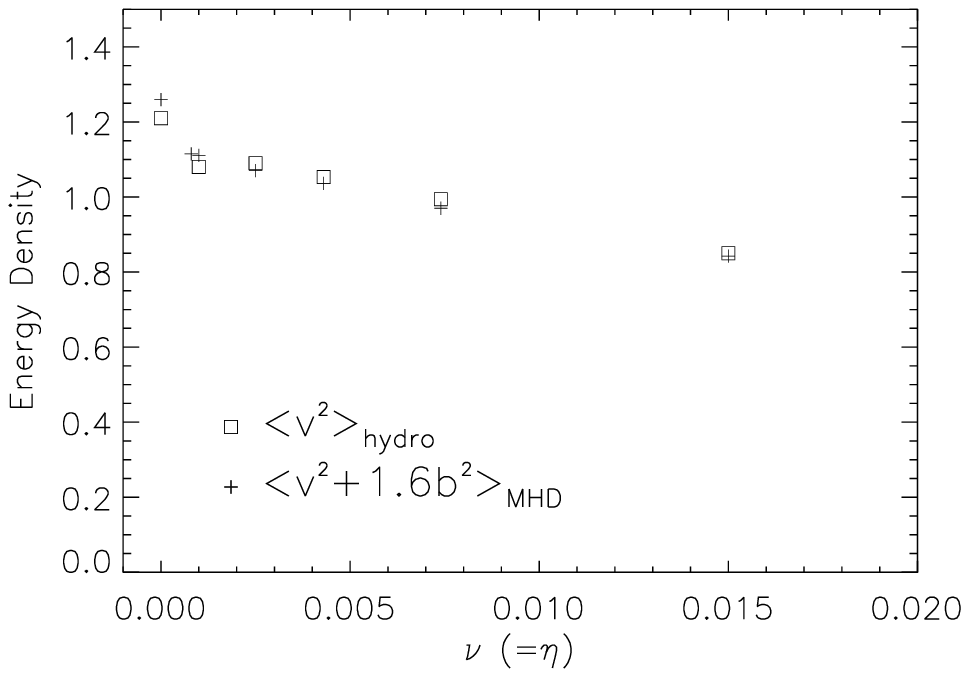}  
\includegraphics[width=.45\textwidth]{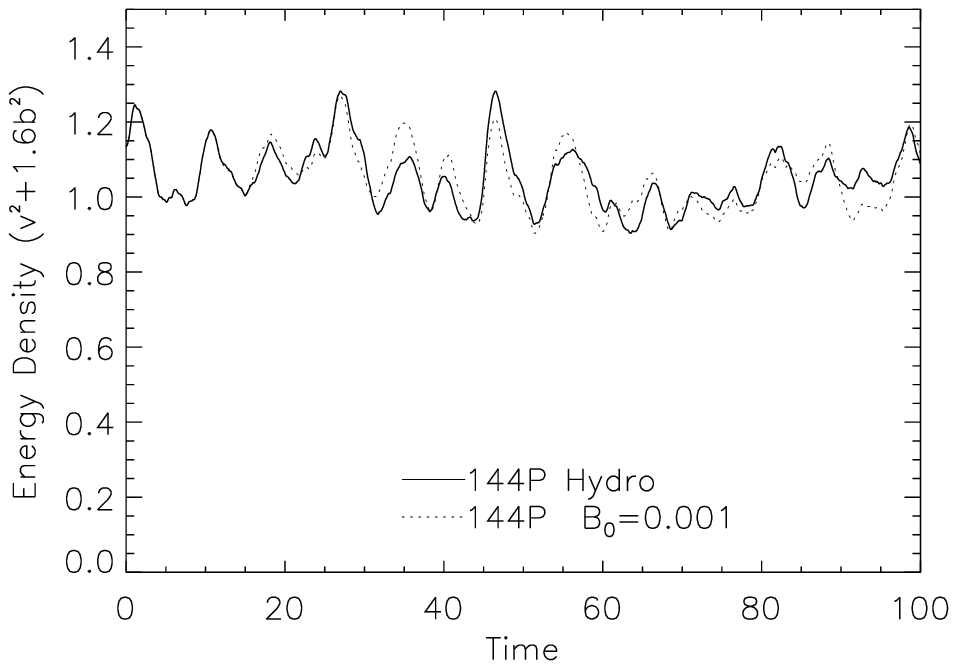}  
\caption{Comparison between MHD and hydrodynamic turbulence.
{\it (Left panel):} Average energy density at the saturation stage.
In MHD cases, $B_0=0.001$ for all simulations. 
The runs that correspond to $\nu \approx 0$ are 256H3-Hydro and
256H3-$B_010^{-3}$.
{\it (Right panel):} The similarity between $v^2_{hydro}$ and
$(v^2+1.6b^2)_{MHD}$ holds true even for time evolution. 
\label{fig:hydromhd}
}
\end{figure*}

We first compare energy densities.
We run hydrodynamic and MHD simulations with the same initial conditions,
except $B_0$, and the same prescribed driving force.
Then we compare energy densities at the saturation stage.
Fig.~\ref{fig:hydromhd} shows the behavior of the energy densities
as a function of $\nu~(=\eta)$.
It is interesting that $<v^2>$ in hydrodynamic turbulence and
$<v^2+1.6b^2>$ in MHD cases scale similarly.
Right panel of the figure shows that the similarity is also true for
individual time basis.

\begin{figure*}[h!t]
\includegraphics[width=.45\textwidth]{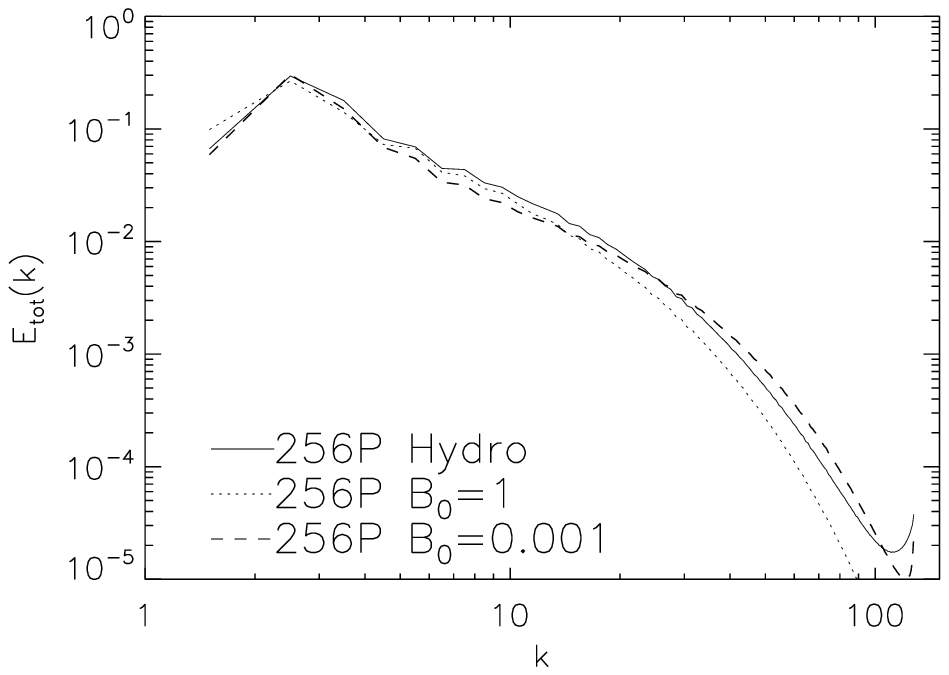} 
\includegraphics[width=.45\textwidth]{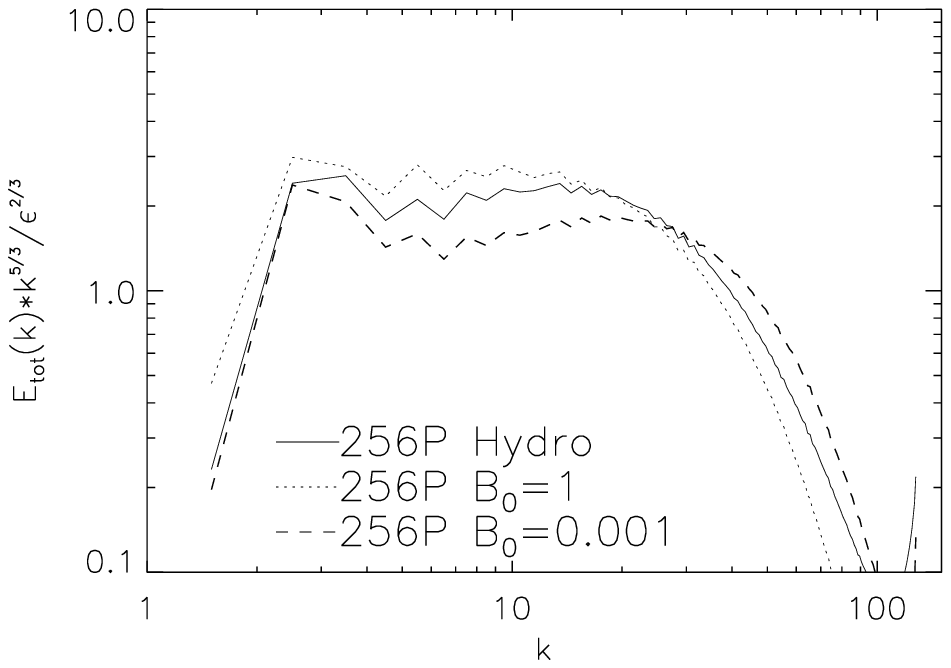} 
\caption{Comparison of energy spectra.
{\it (Left panel):} Total energy spectra for MHD and kinetic energy spectrum for 
hydrodynamic turbulence. Right panel shows that the slope is
compatible with Kolmogorov spectrum.
{\it (Right panel):} Compensated spectra show that, 
when $B_0$ is weak, the total energy spectrum of MHD is lower than 
hydrodynamic one. This is understandable because 
$(v^2+b^2)_{MHD} < (v^2+1.6b^2)_{MHD} \sim v^2_{hydro}$ regardless of
the strength of the mean magnetic field and because
$\epsilon$'s in MHD with $B_0\approx 0$ and 
hydrodynamic cases are nearly same.
However, when $B_0\sim v$, $\epsilon$ reduces and, therefore,
the compensated spectrum is higher than  the hydrodynamic one.
We plot spectra averaged over time.
\label{fig:spect_hdmhd}
 }
\end{figure*}

The claim has a long history that total energy spectrum in MHD turbulence
follows a Kolmogorov spectrum 
(see, for example, Kida, Yanase, \& Mizushima 1991; M\"uller \&
Biskamp 2000).
We plot the energy spectra of hydrodynamic and MHD turbulence
in Fig.~\ref{fig:spect_hdmhd}.
The compensated spectra in right panel of the figure 
shows that the total energy does show a slope
compatible with the Kolmogorov one.
However, the compensated spectrum of MHD turbulence can be higher or lower than
that of hydrodynamic one depending on the mean field strength.
In fact, the vertical location in the right panel of Fig.~\ref{fig:spect_hdmhd} corresponds to the Kolmogorov constant $C_K$:
\begin{equation}
  E(k) = C_K \epsilon^{2/3} k^{-5/3}.
\end{equation}
The Kolmogorov constant for hydrodynamic turbulence is around 1.6 (see, for example,
Yeung \& Zhou 1996).
On the other hand, in MHD case, Biskamp \& M\"uller (2000) reported that
$C_K \sim 2.3$ in their decaying turbulence simulations.
Our results show that the constant can be a function of $B_0$.

\begin{figure*}[h!t]
\includegraphics[width=.45\textwidth]{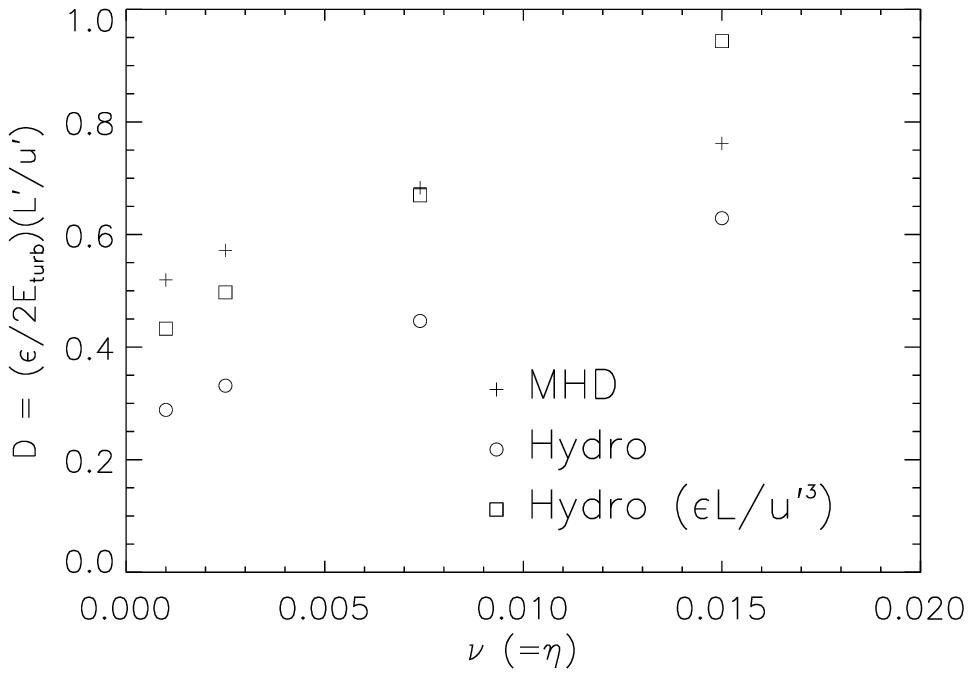}
\caption{Dimensionless energy dissipation coefficient. Both hydrodynamic and
MHD cases are shown.
In MHD cases, $B_0=0.001$ for all simulations. 
 See Eqs. (\ref{eq:diss}) and (\ref{eq:kaneda}) 
for the definition of the coefficient.
\label{fig:diss_hdmhd}
}
\end{figure*}

We compare the dimensionless dissipation coefficient 
(equation [\ref{eq:diss}])
for hydrodynamic and MHD
turbulence in Fig.~\ref{fig:diss_hdmhd}.
We define the coefficient as in equation (\ref{eq:diss}). For MHD cases 
we use $E_{turb}=(v^2+b^2)/2$ and, for hydrodynamic cases, $E_{turb}=v^2/2$. 
In the plot, we also mark the coefficient defined as in 
equation (\ref{eq:kaneda}), where ${u^{\prime}}^2$ is used 
instead of $2E_{turb}$, for hydrodynamic cases (see squares on the Figure).
In hydrodynamic runs, the coefficient is dependent on the value of $\nu$.
However, in MHD runs, the coefficient does not show strong dependence on
$\nu~(=\eta)$.

\section{Discussion}

\subsection{Magnetic reconnection}

Astrophysical magnetic fields are really ubiquitous. 
The origin of the large-scale regular magnetic fields is uncertain.
Nevertheless, the turbulence dynamo discussed in this paper 
provides a simple and efficient way to produce small-scale random fields. 
It only requires the existence of turbulence, which is
as ubiquitous as magnetic fields themselves 
(see Armstrong, Spangler \& Rickett 1995; Lazarian \& Pogosyan 2000;
McKee \& Ostriker 2007). 
Nevertheless, 
small-scale turbulence
dynamo assumes fast magnetic reconnection.

The reconnection is, indeed, fast in the numerical simulation that we deal with here. 
If, in astrophysical fluids, magnetic
reconnection is slow, i.e. happens on time scales longer than 
the dynamical time, both the models of strong turbulence 
and the generation of magnetic field by turbulence are misrepresented by the numerics. 
In this situation we expect the formation of unresolved magnetic knots that should substantially alter the dynamics of the fluid and magnetic fields. 
If, however, reconnection if fast, i.e. happens on the time scales
comparable to the Alfvenic wave propagation time scale,  
then our calculations do represent generation of astrophysical magnetic fields. 
We have ample astrophysical evidence and 
some models of fast reconnection are rather robust. 
For instance, a model of turbulent reconnection in 
Lazarian \& Vishniac (1999; 
see also Vishniac, Lazarian \& Cho 2003; Lazarian, Vishniac \& Cho 2004) 
predicts that magnetic field
change its topology over the eddy turnover time. 
If, however, magnetic reconnection is not a robust process, 
i.e. is fast only in particular, e.g. collisionless environments 
(see Shay et al. 1999), the turbulent generation of magnetic field will differ substantially from one environment to another. 
Future testing of magnetic reconnection should clarify this issue.

\subsection{Turbulence with moderate imposed fields in astrophysical flows}
We have considered the effects of weak mean magnetic fields
($ 0 \leq B_0/\sqrt{4 \pi \rho } \leq v_{rms}$)
on MHD turbulence.
Note that in our simulations $<v^2>=v_{rms}^2 \sim 1$ and effectively we can assume
$\rho=1$ for incompressible runs. In our code units, the factor $4\pi$ disappears.
Therefore, the parameter space we have considered is
$ 0 \leq B_0 \leq 1$.
This work provides missing link between two
extreme limits: $B_0\rightarrow 0$ limit and $B_0/\sqrt{4 \pi \rho} 
\geq v_{rms}$
limit.

The limit of $B_0 \rightarrow 0$ has been studied by many researchers
(Pouquet \& Patterson 1978; Meneguzzi, Frisch, \& Pouquet 1981;
 Kida, Yanase, \& Mizushima 1991; Cho \& Vishniac 2000a).
In this paper,
   we have shown that  there is no difference between turbulence with
no external magnetic field and one with very weak external fields.
In this sense,  the limit of $B_0\rightarrow 0$ is not a special limit.
Cho \& Vishniac (2000a) showed that
energy equipartition between small scale kinetic energy and 
total magnetic energy occurs at a scale about 3 times smaller than
the energy injection scale.
Magnetic spectrum of hyperviscosity simulation peaks at a wavelength about 2-3
time larger the kinetic energy peak.  
These results imply that, when we have a tangled magnetic field with no 
mean magnetic field in a fully
developed turbulence,
the characteristic size, $l_B$, of the magnetic field is about
2-3 times smaller than the energy injection scale of the turbulence.

When $ 0 < B_0/\sqrt{4 \pi \rho } < v_{rms}$,
the turbulence is called {\it super-Alfvenic} in the interstellar medium community.
The intracluster medium may fall in this regime (Ryu et al.~2008).
Although it is still uncertain, the interstellar medium and/or molecular
clouds  in Galaxy
may also fall in this regime (Padoan et al. 2004a; Beck 2001).
Beck (2001) gave an estimate of field strength derived from
many different observations. 
He proposed that the regular component in our galaxy 
is $\sim 4 \mu$G and the total
component $\sim 6\mu$G.
On the other hand,
Heiles \& Crutcher (2007) suggested that the energy ratio
$E_{turb}/E_{mag}$ for cold neutral medium (CNM) is $\sim 1.3$
in our galaxy.
Fig. \ref{fig:v2b2} shows that, when $B_0 > 0.1$, $B^2$ can be
larger than $v^2$ for 256H3 runs.
Therefore, we expect that
the mean field 
in the interstellar medium should be larger than $\sim 0.1$.
Note that, when $B_0$ is larger than $\sim 0.2$, turbulence becomes
 globally anisotropic (Fig.~\ref{fig:ani}).
Therefore, ISM turbulence is either marginally anisotropic 
if $B_0 \sim 0.1$ or anisotropic if $B_0>0.2$.
Turbulence in this case 
will reach saturation level very quickly within 3 or 4 eddy 
turnover times at most.
The correlation length scale of magnetic field will be very close to
the energy injection scale.

\subsection{Model for growth of magnetic field in astrophysical flows}
\label{sect_actualastro}
\begin{figure*}[h!t]
\includegraphics[width=.45\textwidth]{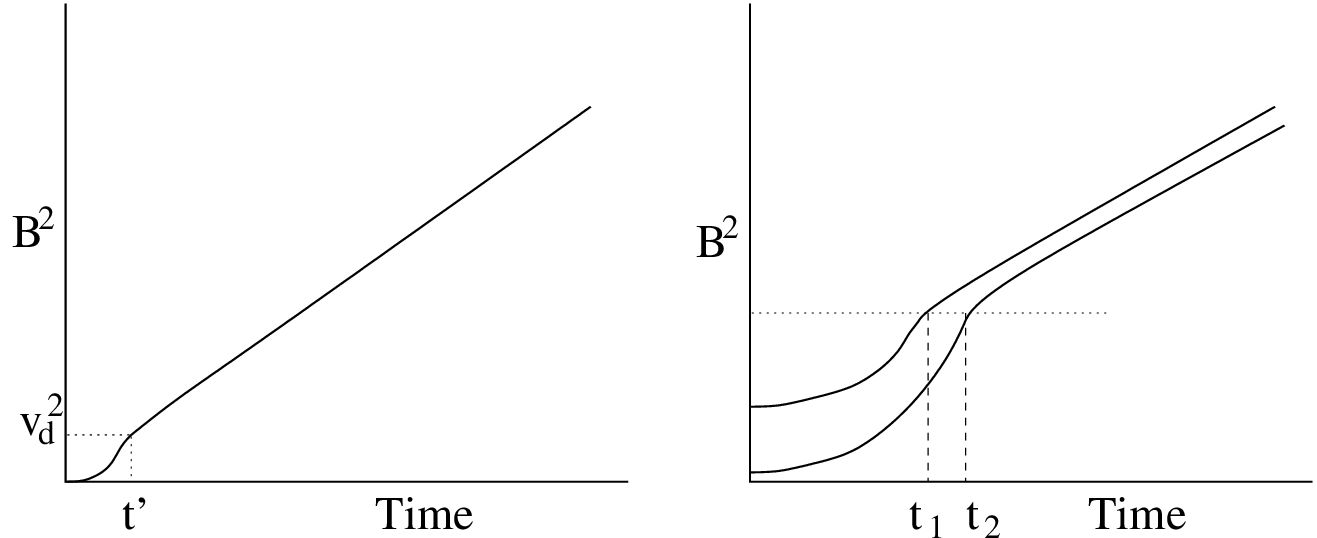}  
\caption{Growth of magnetic field energy in the presence of very weak
mean magnetic field.
 {\it (Left panel):} Initially magnetic energy grows exponentially.
E-folding time in this stage is the eddy turnover time at the
dissipation scale.
  When the magnetic energy reaches equipartition with kinetic energy
at the dissipation scale, exponential stage ends and linear growth
stage begins.
In actual astrophysical fluids with high Reynolds number, 
$t^{\prime}$ and $v_d^2$ are
very small and, therefore, we expect $B^2(t) \propto t$ for most
of the time until the saturation stage.
 {\it (Right panel):} When the mean field strengths are different, the
linear growth stage begins at different times.
When the Reynolds number is high, 
$t_2-t_1$ will be very small.
\label{fig:drawing}
}
\end{figure*}

In numerical simulations, the dissipation scale 
cannot be arbitrarily small. Instead, the dissipation scale is limited by
the numerical resolution. In actual astrophysical fluids,
the dissipation scale will be much smaller than the one we can achieve with
numerical simulations. 
Then what will be the time evolution of the magnetic energy in an actual
astrophysical fluid which has a very high Reynolds number?
Here, we only consider the case that $B_0$ is extremely small.
When numerical resolution or Reynolds number is high, the dissipation scale 
is small and  the eddy turnover time at the dissipation scale is short. 
As a result,
the exponential growth is fast. 
Suppose that the linear growth stage begins at $t=t^{\prime}$.
We plot the situation in left panel of Fig.~\ref{fig:drawing}.
Time evolution of magnetic energy for $t>t^{\prime}$ is
\begin{equation}
   B^2(t) - v_{d}^2 = C( t-t^{\prime} ), \label{eq:b1t} 
\end{equation}
where, $C$ is the slope at the linear growth stage.
The kinetic energy at the dissipation scale, $\sim v_{d}^2$, and
the time $t^{\prime}$ are given by
\begin{equation}
     v_d^2 \sim v^2(l_d/L)^{2/3} \sim v^2 ~(Re)^{-1/2},  \label{eq:vd2}
\end{equation}
and 
\begin{eqnarray}
       t^{\prime} \sim \tau_d \ln (v_d/B_0)^2 \nonumber \\
        \sim t_{eddy} (l_d/L)^{2/3} \ln (Re^{-1/4} v/B_0)^2  \nonumber \\
        \sim t_{eddy} ~(Re)^{-1/2} \ln (Re^{-1/4} v/B_0)^2, \label{eq:tpri}
\end{eqnarray}
where $t_{eddy}$ is the large scale eddy turnover time and
we used the fact $l_d v_d/\nu \sim 1 \sim (Lv/\nu) (l_d/L)(v_d/v)\sim
(Re)~(l_d/L)^{4/3}$. Here $l_d$ is the dissipation scale, 
$v$ the large scale rms velocity,
$v_d$ the velocity at the dissipation scale, 
 $\tau_d$ is proportional to the eddy turnover time at the dissipation scale,
and
$Re=Lv/\nu$ the Reynolds number.
Eqs.~(\ref{eq:vd2}) and (\ref{eq:tpri}) imply that
both 
 $v_{d}^2$ and $t^{\prime}$ go to zero 
when Reynolds number is very large. 
Therefore, we get
\begin{equation}
     B^2(t)  \approx C t   \label{b1real}
\end{equation}
in real astrophysical fluids with very large Reynolds numbers.
When numerical resolution is finite, time evolution of $B^2$
will look like
\begin{equation}
       B^2(t)  \approx C (t - t^{\prime}) 
\end{equation}
because $v_d^2$ is negligible for reasonably high numerical resolutions.

Results in \S\ref{sect:growthB} (see left panel of Fig. \ref{fig:tests})
 show that
 the onset of the linear stage occurs later when the mean field is weak.
Then how large is the delay?
We plot the situation in right panel of Fig.~\ref{fig:drawing}. 
Note that the linear stage begins when energy equipartition is reached
at the dissipation scale, which should be independent of
the strength of the mean field.
When the mean field is strong, the linear stage begins at $t_1$.
When the mean field is weak, the linear stage begins later at $t_2$.
Roughly speaking, the magnetic energy at the exponential growth stage is
\begin{equation}
  B^2(t) \propto B_0^2 \exp(-t/\tau_d).
\end{equation}
Therefore, we have
\begin{equation}
  B_{0,1}^2 \exp(-t_1/\tau_d) = B_{0,2}^2 \exp(-t_2/\tau_d).
\end{equation}
Solving this equation, we get
\begin{eqnarray}
 t_2 - t_1 \propto \tau_d \ln( B_{0,1}/B_{0,2} ) \nonumber  \\
           \propto t_{eddy} ~(Re)^{-1/2} \ln( B_{0,1}/B_{0,2} ).
\end{eqnarray}
Of course, when the Reynolds number is large, $t_2-t_1$ is nearly zero,
which mean that the strength of the mean field does not
matter much in astrophysical fluids.

\section{Conclusion}
We have studied the growth of magnetic field in the presence of weak/moderate
mean magnetic fields.
We have considered only the cases of $\nu=\eta$.
We have found that equipartition between the kinetic and magnetic 
energy densities occurs at a scale somewhat smaller than the kinetic energy peak,
which is consistent with previous results with lower numerical resolution 
 by Cho \& Vishniac (2000a).
We have found that runs with different numerical resolutions and 
simulation parameters show consistent results for
the slope of the linear growth stage.
In real astrophysical flows with very weak mean field and very large Reynolds
number, the growth of magnetic energy will follow
\begin{equation}
      \frac{ B^2(t) }{ 2 E_{turb} } \sim 0.033  
    \frac{ t }{ L/\sqrt{v^2+B^2} }.
\end{equation}

We have also studied the effects of external magnetic fields on MHD turbulence.
We have shown that magnetic energy density increases and kinetic energy
density decreases as the external field becomes stronger.
To be specific, we have shown that, when $B_0 \leq 0.2$,
\begin{eqnarray}
  b^2 - b^{(0)2} \propto vB_0 , \label{conbsq}\\
  v^2 - v^{(0)2} \propto -vB_0, \label{convsq}
\end{eqnarray}
where $b^2$ is the magnetic energy density in the presence of the external field
and $b^{(0)2}$ is the value when the external field is zero.

When the external magnetic field 
is not very strong ($\sim$ less than 0.2 times the rms velocity),
the turbulence remains statistically isotropic, i.e. there
is no apparent anisotropy of order $B_0/v$.
The slope of kinetic spectra at small k's is a function of $B_0$:
when $B_0$ is small, kinetic spectra should be steeper than $k^{-5/3}$ and
those for larger $B_0$ are similar to $k^{-5/3}$.

We have compared MHD and hydrodynamic turbulence.
Spectra of total energy in MHD turbulence are similar to those of 
hydrodynamic turbulence.
The Kolmogorov constant in MHD cases can be higher or lower than that of
hydrodynamic turbulence, depending on the strength of the mean field.

\acknowledgements
Jungyeon Cho's work was supported by the Korea Research Foundation grant
funded by the Korean Government (KRF-2006-331-C00136). 
A. Beresnyak acknowledges the support of the Ice Cube project. A. Lazarian acknowledges the support by the NSF grants AST AST 0507164, as well as by the
NSF Center for Magnetic Self-Organization in Laboratory and Astrophysical 
Plasmas. 
The work of Dongsu Ryu was supported by KOSEF through the grant of
the basic research program R01-2007-000-20196-0.
The works of Jungyeon Cho and Dongsu Ryu was also supported by KICOS through
the grant K20702020016-07E0200-01610 provided by MOST.

\clearpage
\begin{deluxetable}{llllclllllll}  
\footnotesize
\tablecaption{Results of Simulations}
\tablewidth{0pt}
\tablehead{
\colhead{Run \tablenotemark{\dagger}} & \colhead{$N^3$} & \colhead{$\nu=\eta$} & 
\colhead{$B_0^2$} & \colhead{$V^2$} & \colhead{$b^2$~\tablenotemark{\ddagger}} & 
\colhead{$\epsilon$} &
\colhead{$D_M$} & 
\colhead{($t_1,t_2$)\tablenotemark{a}}
}
\startdata
320P-$B_010^{-3}$ & $320^3$ & .0008 & $10^{-6}$ &     .662 &    .283    & .157  &  .110    &    (60,122) \\
 \hline
256H8-$B_00$ & $256^3$ & hyper & 0  &   .616 &  0.455  &   
                       - & - & (60,150) \\
256H8-$B_010^{-6}$ & $256^3$ & hyper & $10^{-12}$ &   .601 &  0.432  &                          - & - & (60,150) \\
256H8-$B_010^{-3}$ & $256^3$ & hyper & $10^{-6}$ & .592  & .443 &                           - & - &  (60,150) \\
256H8-$B_010^{-1}$ & $256^3$ & hyper & $10^{-2}$ &  .495   &  0.566  &                            - & - &  (30,131)\\
\hline
256H3-$B_010^{-3}$ & $256^3$ & hyper & $10^{-6}$ & .613 &  0.404 &                        - & - & (60,150)\\
256H3-$B_010^{-2}$ & $256^3$ & hyper & $10^{-4}$ & .633  & .388  &                        - & - & (60,150)\\
256H3-$B_010^{-1.5}$ & $256^3$ & hyper & $10^{-3}$ &.606 & .417 &                        - & - & (60,150)\\
256H3-$B_010^{-1}$ & $256^3$ & hyper & $10^{-2}$ & .504 &  0.508 &                      - & - & (30,90) \\
256H3-$B_06^{-1}$ & $256^3$ & hyper & 1/36 & .453 &  .584 &                        - & - & (30,46) \\
256H3-$B_010^{-0.5}$ & $256^3$ & hyper & 1/10 & .423 &  .618 &                        - & - & (30,43) \\
256H3-$B_01$ & $256^3$ & hyper & 1 &  .603 & .449  &                        - & - & (15,45) \\
\hline
256P-$B_010^{-3}$ & $256^3$ & 0.001 & $10^{-6}$ &  .711  &  .250 &  .159 &  .109  &    (60,150)\\
256P-$B_010^{-2}$ & $256^3$ & 0.001 & $10^{-4}$ & .672  & .275 &  .158 &  .110  &    (60,135)\\
256P-$B_010^{-1.5}$ & $256^3$ & 0.001 & $10^{-3}$ & .639 & .285 & .155 &  .107  &     (60,120)\\
256P-$B_010^{-1}$ & $256^3$ & 0.001 & $10^{-2}$ & .559 &  .370 &  .157 &  .108  &    (30,60)\\
256P-$B_05^{-1}$ & $256^3$ & 0.001 & $1/25$ &  .467  &  .460  & .145 & .0970 &    (30,90)\\
256P-$B_010^{-0.5}$ & $256^3$ & 0.001 & $1/10$ & .437 & .500 &   .136 &  .0884 &    (30,90)\\
256P-$B_00.8$ & $256^3$ & 0.001 & .64 &  .512  &   .458 &  .109 & .0609 &    (15,45)\\
256P-$B_01$ & $256^3$ & 0.001 & 1 &  .599 &  .403 & .101 & .0540 &    (15,34)\\
\hline
144H8-$B_010^{-3.5}$&$144^3$&hyper& $10^{-7}$ &.649 & .420& .161 & - & 
(60,240)\\
144H8-$B_010^{-1.5}$&$144^3$&hyper& $10^{-3}$ &.632 & .438& .163 & - & 
(60,220)\\
144H8-$B_016^{-1}$&$144^3$&hyper& 1/256       &.576 & .487& .158 & - & 
(60,220)\\
144H8-$B_010^{-1}$&$144^3$&hyper& 1/100       &.526 & .515& .150 & - & 
(30,90)\\
144H8-$B_010^{-0.5}$&$144^3$&hyper& 1/10      &.458 & .613& .133& - & 
(30,90)\\  \hline
144P-$B_010^{-3}  $&$144^3$&.0025 & $10^{-6}$  &.728 &.214 &.165 &.102 &    (60,350)\\
144P-$B_010^{-2}  $&$144^3$&.0025 & $10^{-4}$  &.743 &.210 &.166 &.102 &
(70,350)\\
144P-$B_010^{-1.5}$&$144^3$&.0025 & $10^{-3}$  &.709 &.227 &.164 &.103 &
(70,350)\\
144P-$B_010^{-1}  $&$144^3$&.0025 &     1/100  &.589 &.312 &.162 &.107 &
(70,350)\\
144P-$B_010^{-0.5}$&$144^3$&.0025 &     1/10   &.450 &.453 &.147 &.095 &
(70,350)\\
144P-$B_01        $&$144^3$&.0025 &     1      &.570 &.346 &.092 &.045 &
(70,350)\\  \hline
96P-$B_00$ &$96^3$& .0043 & 0                & .766 & .165 & .164& .086& 
(200,500)\\
96P-$B_010^{-3.5}$&$96^3$& .0043 & $10^{-7}$ & .761 & .172 & .166& .088& 
(200,500)\\
96P-$B_010^{-2}$&$96^3$& .0043 & $10^{-4}$   & .765 & .168 &.169 & .090& 
(200,500)\\
96P-$B_010^{-1}$ &$96^3$& .0043 &  1/100     & .608 & .268 & .164& .100& 
(200,400)\\ 
96P-$B_010^{-0.5}$ &$96^3$& .0043 &   1/10   & .447 & .404 & .149 & .095& 
(200,400) \\  \hline
64P1-$B_00$ &$64^3$& .0074 &  0                & .784 & .115 & .165& .064& 
(300,800) \\
64P1-$B_010^{-3.5}$ &$64^3$& .0074 & $10^{-7}$ & .786 & .113 & .166& 
.064& (300,800) \\
64P1-$B_010^{-2}$ &$64^3$& .0074 &  $10^{-4}$  & .778 & .119 &.166 & 
.066& (300,800) \\
64P1-$B_010^{-1.5}$ &$64^3$& .0074 & $10^{-3}$ & .743 & .137 & .164& .071& 
(300,800) \\
64P1-$B_010^{-1}$ &$64^3$& .0074 &   1/100     & .617 & .221 &.166 & .089& 
(300,800) \\
64P1-$B_010^{-0.5}$ &$64^3$& .0074 &   1/10    & .441 & .350 &.153 & .093& 
(300,500) \\
64P1-$B_01$         &$64^3$& .0074 &   1       & .519 & .278 &.096 & .044&
(300,800)
\\  \hline
64P2-$B_00$ &$64^3$& .015 & $0  $              & .806 & .0215 &  .164&  .014& 
(300,750) \\
64P2-$B_010^{-3.5}$ &$64^3$& .015 & $10^{-7}$  & .808 & .0201 & .164 &  .013& 
(300,750) \\
64P2-$B_010^{-2}$ &$64^3$& .015 & $10^{-4}$    & .781 & .0364 & .165 &  .022& 
(300,750) \\
64P2-$B_010^{-1.5}$ &$64^3$& .015 & $10^{-3}$  & .730 & .0660 & .166 &  .037& 
(300,750) \\
64P2-$B_010^{-1}$ &$64^3$& .015 &  1/100       & .592 & .145  & .166 &  .065& 
(300,750) \\ 
64P2-$B_010^{-0.5}$ &$64^3$& .015 &  1/10      & .413 & .269  & .155 &  .085& 
(300,750) \\
64P2-$B_01$         &$64^3$& .015 &  1         & .409 & .210  & .095 &  .041&
(300,750)
\\  \hline
256H8-hydro &$256^3$ & hyper & -  & 1.225 & - & .168 & -& (15,30)\\
256H3-hydro &$256^3$ & hyper & -  & 1.207 & - & .165 & -& (15,28)\\
256P-hydro  &$256^3$ & .001  & -  & 1.080 & - & .150 & -& (15,75)\\
144H8-hydro &$144^3$ & hyper & -  & 1.118 & - & -    & -& (8,18)\\
144P-hydro  &$144^3$ & .0025 & -  & 1.090 & - & .164 & -& (60,350)\\
96P-hydro   &$96^3$  & .0043 & -  & 1.053 & - & .160 & -& (50,165)\\
64P1-hydro  &$64^3$  & .0074 & -  & .994  & - & .165 & -& (300,800)\\
64P2-hydro  &$64^3$  & .015  & -  & .850  & - & .163 & -& (300,750)\\
             \hline
\enddata
\tablenotetext{\dagger}{Only selected runs are shown in this table.}
\tablenotetext{\ddagger}{$B^2=B_0^2+b^2$}
\tablenotetext{a}{Time interval used for averaging physical quantities}
             
\label{table_2}
\end{deluxetable}

\end{document}